\documentclass[prb,floatfix,aps,twocolumn,showpacs]{revtex4-1}
\usepackage[dvips]{graphics}
\usepackage{graphicx}
\usepackage[utf8]{inputenc}
\usepackage{amssymb}
\usepackage{epstopdf}
\usepackage{bm}
\usepackage{dcolumn}
\usepackage{epsfig}
\usepackage{latexsym}
\usepackage{amsmath}
\usepackage{color}
\usepackage{array}
\usepackage{enumerate,dsfont}

\def\XXint#1#2#3{{\setbox0=\hbox{$#1{#2#3}{\int}$ }
\vcenter{\hbox{$#2#3$ }}\kern-.5\wd0}}

\newcommand{\ket}[1]{\left |\mbox{$#1$}\right\rangle}
\newcommand{\bra}[1]{\left\langle\mbox{$#1$}\right |}

\newcommand{\arccsc}{\mathrm{acsc}}
\newcommand{\angstrom}{\textup{\AA}}

\newcommand{\sub}[1]{_{\rm #1}}

\newcommand{\cre}[1]{#1^{\dagger}}

\begin{document}
\title{Long-Range Interaction of Singlet-Triplet Qubits via Ferromagnets}
\author{Luka Trifunovic, Fabio L. Pedrocchi, and Daniel Loss}
\affiliation{Department of Physics, University of Basel, Klingelbergstrasse 82, CH-4056 Basel, Switzerland}

\begin{abstract}
We propose a mechanism of a long-range coherent interaction between two
singlet-triplet qubits dipolarly coupled to a dogbone-shaped ferromagnet. An
effective qubit-qubit interaction Hamiltonian is derived and the coupling
strength is estimated. Furthermore we derive the effective coupling between two
spin-$1/2$ qubits that are coupled via dipolar interaction to the ferromagnet
and that lie at arbitrary positions and deduce the optimal positioning. We
consider hybrid systems consisting of spin-$1/2$ and ST qubits and derive the
effective Hamiltonian for this case. We then show that operation times  vary
between 1MHz and 100MHz and give explicit estimates for  GaAs, Silicon, and
NV-center based spin qubits. Finally, we explicitly construct the required
sequences to implement a CNOT gate. The resulting quantum computing
architecture retains all the single qubit gates and measurement aspects of
earlier approaches, but allows qubit spacing at distances of order 1$\,\mu$m
for two-qubit gates, achievable with current semiconductor technology. 
\end{abstract}

\maketitle
\section{Introduction}
At the heart of quantum computation lies the ability to generate, measure, and
control entanglement between qubits. This is a difficult task because the qubit
state is rapidly destroyed by the coupling to the environment it is living in.
Hence, a qubit system appropriate for quantum computing must be sufficiently
large for it to be controllable by experimentalist but must contain as few
degrees of freedom as possible that couple to the environment. One of the most
successful candidates for encoding a qubit is an electron spin localized in a
semiconductor quantum dot,  gate-defined or self-assembled, or a singlet-triplet
qubit with two electrons in a double quantum
well.~\cite{kloeffel_prospects_2013,levy_universal_2002} These natural two-level
systems are very long-lived (relaxation time $T_1\sim1s$, see
Ref.~\onlinecite{amasha_electrical_2008}, and decoherence time $T_{2}> 200 \mu
s$, see Ref.~\onlinecite{Bluhm_Dephasing_2011}), they can be controlled
efficiently by both electric and magnetic
fields,~\cite{petta_coherent_2005,koppens_spin_2008,brunner_two_2011} and,
eventually, may  be scaled into a large network. It has been experimentally
demonstrated that qubit-qubit couplings can be generated and controlled
efficiently for these systems.~\cite{shulman_demonstration_2012} However the
separation between the quantum dots needs to be small ($\sim 100nm$)  and this
renders their scaling to a very large number of qubits a perplexed task. Indeed,
the physical implementation of quantum dot networks requires some space between
the qubits for the different physical auxiliary components (metallic gates,
etc.). It is therefore important to find a way to couple qubits over
sufficiently large distances (micrometer scale) to satisfy the space constraint.

Another type of promising two-level systems are silicon-based spin qubits. They
are composed of nuclear (electron) spin of phosphorus atoms in a silicon
nanostructure. It has recently been shown that very long decoherence times
$T_{2}\approx60ms$~\cite{pla_high_2013} ($T_{2}\approx 200 \mu
s$~\cite{pla_single-atom_2012}) and high fidelity single qubit gates and
readout are  experimentally achievable. However, the realization of such qubits
is subject to randomness of the phosphorus atom position in silicon. Hence,
their location is a priori unknown. If two randomly chosen qubits are
well-separated from each other, they will hardly interact. On the contrary,
when they lie close to each other, it will be difficult to turn-off the
interaction since they will not be isolated from each other. In this context it
is especially important to be able to couple two qubits over relatively large
distances by putting some coupler between them. The aim of our work is to show
that this is possible by putting a micrometer-sized ferromagnet between the
qubits. We point out that our analysis is general and does not depend on the
precise nature of the qubits that need to be coupled as long as they interact
dipolarly with the ferromagnet. Hence we think that our analysis is applicable
to a variety of other spin qubits systems such as N-V centers in
diamond.~\cite{dobrovitski_quantum_2013}

There have been various other propsals over the last years in order to couple
spin qubits over large distances. Among them we mention here coupling through a
superconductor,~\cite{choi_spin-dependent_2000,leijnse_coupling_2013} microwave
cavities,~\cite{trif_spin_2008,Wallraff_Strong_2004} two-dimensional electron gas
(RKKY),~\cite{rikitake_decoherence_2005} and floating
gates.~\cite{flensberg_bends_2010,trifunovic_long-distance_2012,flindt_spin-orbit_2006,trif_spin-spin_2007}

\begin{figure}
  \includegraphics[width=\columnwidth]{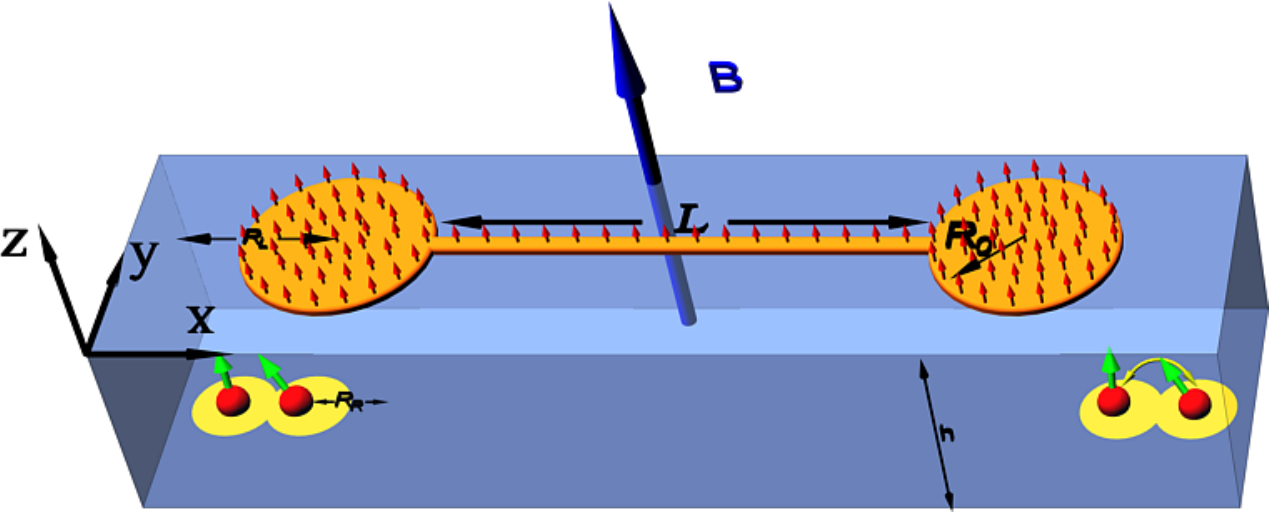}
  \caption{Model system consisting of two identical double-QDs in the $xy$-plane
  and the dogbone-shaped coupler. The dogbone coupler consists of two
  ferromagnetic disks of radius $R_0$ connected by a thin ferromagnetic wire of
  length $L$. Each double-QD can accommodate one (two) electrons, defining the
  spin-$1/2$ (ST-) qubit. Absence of tunneling between the separate double-QD is
  assumed.  Here $R_{\rm L}$ ($R_{\rm R}$) is the in-plane distance between the
  left (right) well and the corresponding disk center, while $h$ is vertical
  distance between the QD and the gate. The red arrow on top of the ferromagnet
  denote the orientation of its magnetization which is assumed to be monodomain.}
  \label{fig:setup-fig}
\end{figure}

In this work, we propose and study a system that allows for coherent coupling
between ST-qubits as well as between spin-$1/2$ qubits over distances of about
one micrometer. The coupler is a ferromagnet composed of two disks separated by
a thin quasi-1D region, see Fig.~\ref{fig:setup-fig}. The qubits are coupled to
the ferromagnet via dipolar interaction and they are positioned in the vicinity
of each disk. The relevant quantity of the coupler, describing the effective
coupling between the distant qubits, is its spin-spin susceptibility---a slowly
spatially decaying real part of the susceptibility is required in order to
mediate interactions over long distances. Additionally, in order to have
\textit{coherent} coupling, the imaginary part of the susceptibility should be
sufficiently small. The spatial decay of the spin susceptibility depends
strongly on the dimensionality of the ferromagnet---it is longer ranged in lower
dimensions. The dogbone shape of the coupler considered here is thus optimal: it
allows for strong coupling to the ferromagnet because many spins of each disk
lie close to the qubits, while the coherent interaction between them is mediated
by the quasi-1D channel. Actually the statement on the optimal shape of the
\textit{coupler} is quite general if we consider a realistic interaction between
the qubit and the coupler, i.e., a Coulomb~\cite{trifunovic_long-distance_2012}
or a dipolar one as herein.

Likewise, we derive a Hamiltonian for the effective interaction between the
distant qubits positioned arbitrarily with respect to each disk of the dogbone
and determine what is the optimal position for the coupling to be strongest. For
ST-qubit the optimal coupling is obtain if one quantum well is positioned
directly below the disk center and the other one below the edge of the disk,
while for the spin-$1/2$ qubits it is optimal to place them at the edge of the
disk. Similar conclusions about the optimal positioning of the qubits with
respect to the coupler were previously obtained for the case of electrostatic
interaction.~\cite{trifunovic_long-distance_2012}

In the most favorable scenarios described above the coupling strength of
$10^{-2} \mu{\rm eV}$ are achieved for ST-qubits as well as for spin-$1/2$
qubits. The clock speed for such coupling schemes thus varies between 1MHz and
100MHz. We summarize in Tables~\ref{tab:1} and \ref{tab:2} of
Section~\ref{sec:Operation_Times} the coupling strengths and corresponding
operation times achievable with our scheme.

In order to be useful for quantum computation, one needs to be able to turn on
and off the coupling between the qubits. This is efficiently achieved by
putting the qubit splitting off-resonance with the internal splitting of the
ferromagnet. As we argue below, a modification in the qubit splitting of about
one percent of $\Delta_F$ is enough to interrupt the interaction between the
qubits. Finally we derive for both qubit systems the sequence to implement the
entangling gates CNOT (and iSWAP) that can be achieved with a gate fidelity
exceeding 99.9\%. The additional decoherence effects induced solely by the
coupling to the ferromagnet are negligible for sufficiently low temperatures
($T\lesssim0.1K$).\cite{2013arXiv1302.4017T} We then obtain error
thresholds---defined as the ratio between the two-qubit gate operation time to
the decoherence time---of about $10^{-4}$ for ST-qubits as well as for spin-1/2
qubits and this is good enough to implement the surface code error correction
in such setups. This quantum computing architecture thus retains all the single
qubit gates and measurement aspects of earlier approaches, but allows qubit
spacing at distances of order 1$\,\mu$m for two-qubit gates, achievable with
current semiconductor technology.

The paper is organized as follows. In Sections~\ref{sec:ferromagnet} and
\ref{sec:STA} we introduce the ferromagnet and ST-qubit Hamiltonian,
respectively. In Sec.~\ref{sec:STB} we derive the effective dipolar coupling
between the ST-qubit and the ferromagnet. In Sec.~\ref{sec:STC} we make use
of a perturbative Scrhieffer-Wolff transformation to derive the effective
coupling between the two ST-qubits that is mediated by the ferromagnet. We
determine the optimal position of the qubits relative to the disks of the
dogbone. In Sec.~\ref{sec:CNOT} we construct the sequence to implement the CNOT
(and iSWAP) gate and calculate the corresponding fidelity of the sequence. In
Sec.~\ref{sec:onehalf} we study the coupling between two spin-$1/2$ qubits
positioned at arbitrary location with respect to the adjacent disk of the
dogbone-shaped ferromagnet. We derive an effective Hamiltonian for the
interaction of the two spin-$1/2$ qubits mediated by the ferromagnet and
determine the optimal position of the qubits. In Sec.~\ref{sec:onehalfA} we
derive the sequence to implement the CNOT (and iSWAP) gate. In
Sec.~\ref{sec:Hybrid}, we show that spin-$1/2$ and ST qubits can be
cross-coupled leading to hybrid qubits and we derive the effective Hamiltonian
for this case. In Sec.~\ref{sec:validity}, we discuss the range of validity of
our effective theory. The \textit{on/off} switching mechanisms of the
qubit-qubit coupling are discussed in Sec.~\ref{sec:switching}. In
Sec.~\ref{sec:Operation_Times}, we present a table with the effective coupling
strengths and operation times achievable in our setup for four experimentally
relevant systems, namely GaAs spin-$1/2$ quantum dots, GaAs singlet-triplet
quantum dots, silicon-based qubits, and N-V centers. Finally,
Sec.~\ref{sec:conclusions} contains our final remarks and the Appendices
additional details on the models and derivations.
\section{Ferromagnet}\label{sec:ferromagnet}
We denote by $\bm S_{ \bf r}$ the spins (of size $S$) of the ferromagnet at site
${\bf r}$ on a cubic lattice and $\bm\sigma_i$ stands for the spin-$1/2$ qubit
spins. The ferromagnet Hamiltonian we consider is of the following form
\begin{equation}
  \label{eq:ferromagnet}
  H_{F}=-J\sum_{\langle {\bf r},{\bf r'}\rangle}{\bm S}_{\bf r}\cdot{\bm S}_{
  \bf r'}+\Delta_F\sum_{{\bf r}}S^z_{\bf r},
\end{equation}
with $J>0$ and $\Delta_F=\mu B$, where $B$ is externally applied magnetic field
(see Fig.~\ref{fig:setup-fig}) and $\mu$ is the magnetic moment of the
ferromagnet spin. The above Hamiltonian is the three-dimensional (3D) Heisenberg
model with the sum restricted to nearest-neighbor sites $\langle {\bf r},{\bf
r'}\rangle$. The ferromagnet is assumed to be monodomain and below the Curie
temperature with the magnetization pointing along the $z$-direction.

We would like to stress at this point that even though herein we analyze a
specific model for the ferromagnet (Heisenberg model), all our conclusions rely
only on the generic features of the ferromagnet susceptibility, i.e., its
long-range nature. Furthermore, the gap in the magnon spectrum can originate
also from anisotropy. The presence of the gap is an important feature since it
suppresses the fluctuations, albeit the susceptibility is cut-off after some
characteristic length given by the gap and the frequency at which the
ferromagnet is probed.

\section{Coupling between ST-qubits}\label{sec:ST}
The Hamiltonian we consider is of the following form
\begin{equation}
  \label{eq:Hamiltonian}
  H=H_F+H_\tau+H_I,
\end{equation}
where $H_\tau$ is Hamiltonian of the two
ST-qubits~\cite{levy_universal_2002,klinovaja_exchange_2012} and $H_I$ is the
dipolar coupling between the ferromagnet and the ST-qubits (see below).

\subsection{Singlet-Triplet qubit Hamiltonian}\label{sec:STA} A
Singlet-Triplet (ST) qubit is a system that consists of two electrons confined
in a double quantum well. Herein we assume that the wells are steep enough so
that we can consider only one lowest orbital level of each well. Following
Ref.~\onlinecite{stepanenko_singlet_2011}, we consider also the spin space of
the two electrons and write down the total of six basis states
\begin{align}  
|(2,0) S \rangle &= \cre{c}_{{\rm L} \uparrow} \cre{c}_{{\rm L}
  \downarrow} | 0 \rangle,\nonumber
\\
|(0,2) S \rangle &= \cre{c}_{{\rm R} \uparrow} \cre{c}_{{\rm R}
  \downarrow} | 0 \rangle,\nonumber
\\
|(1,1) S \rangle &= \frac{1}{\sqrt{2}} \left( \cre{c}_{{\rm L} \uparrow}
\cre{c}_{{\rm R} \downarrow} - \cre{c}_{{\rm L} \downarrow} \cre{c}_{{\rm
    R} \uparrow} \right) | 0 \rangle,\nonumber
\\
  \label{eq:statesdef}
|T_+ \rangle &= \cre{c}_{{\rm L} \uparrow} \cre{c}_{{\rm R} \uparrow} | 0 \rangle,
\\
|T_0 \rangle &= \frac{1}{\sqrt{2}} \left( \cre{c}_{{\rm L}
  \uparrow}\cre{c}_{{\rm R} \downarrow} + \cre{c}_{{\rm L} \downarrow}
\cre{c}_{{\rm R} \uparrow} \right) | 0 \rangle,\nonumber
\\
|T_- \rangle &= \cre{c}_{{\rm L} \downarrow} \cre{c}_{{\rm R} \downarrow}
| 0 \rangle,\nonumber
\end{align}
where $\cre{c}_{\rm L}$ ($\cre{c}_{\rm R}$) creates an electron in the Wannier
state $\Phi_{\rm L}$ ($\Phi_{\rm R}$). The Wannier states are $\Phi\sub{\rm L,R}
= \frac{1}{\sqrt{1-2sg+g^2}}\left( \varphi\sub{1,2} - g
\varphi\sub{2,1}\right)$, where $s=\langle \varphi\sub{1}|\varphi\sub{2}\rangle
=  \exp[- \left( a / a\sub{B} \right) ^2]$ is the overlap of the harmonic
oscillator ground state wave functions of the two wells, $a\sub{B}=
\sqrt{\hbar/m\omega_0}$ is the Bohr radius of a single quantum dot, $\hbar
\omega_0$ is the single-particle level spacing, and $2a=l$ is the interdot
distance. The mixing factor of the Wannier states is $g=
({1-\sqrt{1-s^2}})/{s}$. Using these six basis states we can represent the
Hamiltonian of the ST-qubits
\begin{equation}
  \label{eq:hspinindependent}
  H_0 = \left(
  \begin{array}{cc}
    H\sub{SS}
    &
    0
    \\
    0
    &
    H\sub{TT,0}
  \end{array}
  \right).
\end{equation}
In writing the above equation we have neglected the spin-orbit interaction
(SOI), thus there are no matrix elements coupling the singlet and triplet
blocks. The effect of SOI in ST-qubit was studied in
Ref.~\onlinecite{stepanenko_singlet_2011} and no major influence on the qubit
spectra was found. 

The two qubit states are $\ket{T_0}$ and, in the absence of SOI, the linear
combination of the singlet states $\ket{S}=\alpha \ket{(2,0)S}+\beta
\ket{(1,1)S}+\gamma\ket{(0,2)S}$, where the coefficients $\alpha,\beta,\gamma$
depend on the detuning $\varepsilon$ between the two quantum wells. In
particular, when $\varepsilon=0$ we have $\ket{S}=\ket{(1,1)S}$. In what
follows, we always consider Hamiltonians only in the qubit subspace, thus the
Hamiltonian of two ST-qubits reads

\begin{equation}
  H_\tau=-\frac{\Delta}{2}\sum_{i=1,2}\tau_i^z,
  \label{eq:qubitH}
\end{equation}
where $\tau^{x,y,z}$ are the Pauli matrices acting in the space spanned by vectors
$\{\ket{S},\ket{T_0}\}$ and $\Delta$ is the ST-qubit splitting.
\subsection{Dipolar coupling to ST-qubit}\label{sec:STB}
In this section we derive the dipolar coupling between the ferromagnet and the
ST-qubit. To this end we first project the Zeeman coupling to the ST-qubit
system on the two-dimensional qubit subspace
\begin{equation}
  H_{\rm Z}=g^{*}\mu_{B}\left(\bm B_{\rm L}\cdot\bm S_{\rm L}+\bm B_{\rm R}\cdot\bm S_{\rm R}\right)\,,
  \label{eq:Zeeman}
\end{equation}
where $\bm B_{\rm L}$ ($\bm B_{\rm R}$) is the magnetic field in the left
(right) quantum well, $S_{\rm L,R}^i=(\sigma^i)_{ss^\prime}c^\dagger_{ {\rm
L,R},s}c_{ {\rm L,R},s^\prime}$ and $g^{*}$ is the effective Land\'e factor. After projecting on the qubit space we obtain
\begin{equation}
  H_{\rm Z}=g^{*}\mu_{B}(B_{\rm L}^z-B_{\rm R}^z)\tau^x.
  \label{eq:Zeemanproj}
\end{equation}
With this result we are ready to write down the ferromagnet/ST-qubit interaction
Hamiltonian
\begin{equation}
  H_I=\sum_{i=1,2}g^{*}\mu_{0}\mu_{B}\left( \hat B_{\rm L}^z(i)-\hat B_{\rm R}^z(i) \right)\tau_i^x,
  \label{eq:HI}
\end{equation}
where index $i$ enumerates ST-qubits, and the magnetic field from the ferromagnet
can be express through the integral over the ferromagnet
\begin{align}
    \label{eq:dipolarB}
    \hat B_{\rm L,R}^z(i)=&\frac{\mu_0\mu}{4\pi a^3}\int d\bm r(i)_{\rm L,R}
    \frac{1}{r(i)_{\rm L,R}^3}\times\\
    &\left( S^z_{\bm r(i)_{\rm L,R}}-\frac{3(\bm S_{\bm
    r(i)_{\rm L,R}}\cdot\bm r(i)_{\rm L,R})r(i)_{\rm L,R}^z}{r(i)_{\rm L,R}^2}
    \right),\nonumber
\end{align}
where the coordinate system for $\bm r(i)_{\rm L}$ ($\bm r(i)_{\rm R}$) is
positioned in left (right) quantum well of the $i$-th qubit.
\subsection{Effective coupling between two ST-qubits}\label{sec:STC}
\label{sec:effcoup}
Given the total Hamitonian, Eq.~(\ref{eq:Hamiltonian}), we can easily derive the
effective qubit-qubit coupling with help of Schrieffer-Wolff transformation
\begin{align}
  H_{\rm eff}=H_\tau-\lim_{\nu\rightarrow0^{+}}\frac{i}{2}\int_0^\infty dt e^{-\nu t}
  \left[H_I(t),H_I\right],
  \label{eq:Heffdef}
\end{align}
with $H_I(t)=e^{iH_\tau t}H_I e^{-iH_\tau t}$.

We assume that the radius of the two disks is much smaller than the distance
between their centers ($R_0\ll L$). Within this assumption we can take for the
susceptibility between two points at opposite disks the same as the 1D
susceptibility. Next we take only on-resonance susceptibility and make use of
the expression $\tau^x(\Delta)=\frac{1}{2}(\tau^x+i\tau^y)$, where
$\tau^x(\omega)$ is the Fourier transform of $\tau^x(t)=e^{iH_\tau t}\tau^x
e^{-iH_\tau t}$. We define the transverse susceptibility in the standard way
\begin{equation}
  \chi_{\perp}(\omega,{\bf r}_{i}-{\bf r}_{j})=
  -i\lim\limits_{\eta\rightarrow0^{+}}\int_{0}^{\infty}dt e^{(-i\omega-\eta)
  t}[S_{{ \bf r}_i}^{+}(t),S_{{ \bf r}_j}^{-}].
  \label{eq:susceptibility}
\end{equation}
The longitudinal susceptibility, defined via
\begin{equation}
  \chi_\parallel(\omega,{\bf r}_{i}-{\bf r}_{j})=
  -i\lim\limits_{\eta\rightarrow0^{+}}\int_{0}^{\infty}dt e^{(-i\omega-\eta)
  t}[S_{{ \bf r}_i}^{z}(t),S_{{ \bf r}_j}^{z}],
  \label{eq:susceptibilitylongt}
\end{equation}
can be neglected compared to the transverse one because the former is smaller by
a factor $1/S$ and is proportional to the magnon occupation number, see Eq.~\ref{eq:chi_par}. Therefore the longitudinal susceptibility vanishes at zero temperature, while the is not the case for the transverse susceptibility. We arrive finally at the following
expression
\begin{equation}
  H_{\rm
  eff}=H_\tau+\frac{9}{4}\mathcal{B}\chi^{\rm 1D}_\perp(\Delta,L)\tau_1^x\tau^x_2,
  \label{eq:Heff}
\end{equation}
where $\mathcal{B}=(\mu_0 \mu)^2 (g^{*}\mu_{B})^{2}(A_L^1-A_R^1)(A_L^2-A_R^2)/16\pi^2a^6$,
$\chi_\perp^{\rm 1D}$ is given in Eq.~(\ref{eq:chi_1D_r}) and
\begin{equation}
  A_{\rm L,R}^i=\int d\bm r(i)_{\rm L,R}
  \frac{r(i)^-_{\rm L,R}r(i)_{\rm L,R}^z}{r(i)_{\rm L,R}^5}.
  \label{eq:constA}
\end{equation}
\begin{figure}[h]
  \centering\includegraphics[width=0.9\columnwidth]{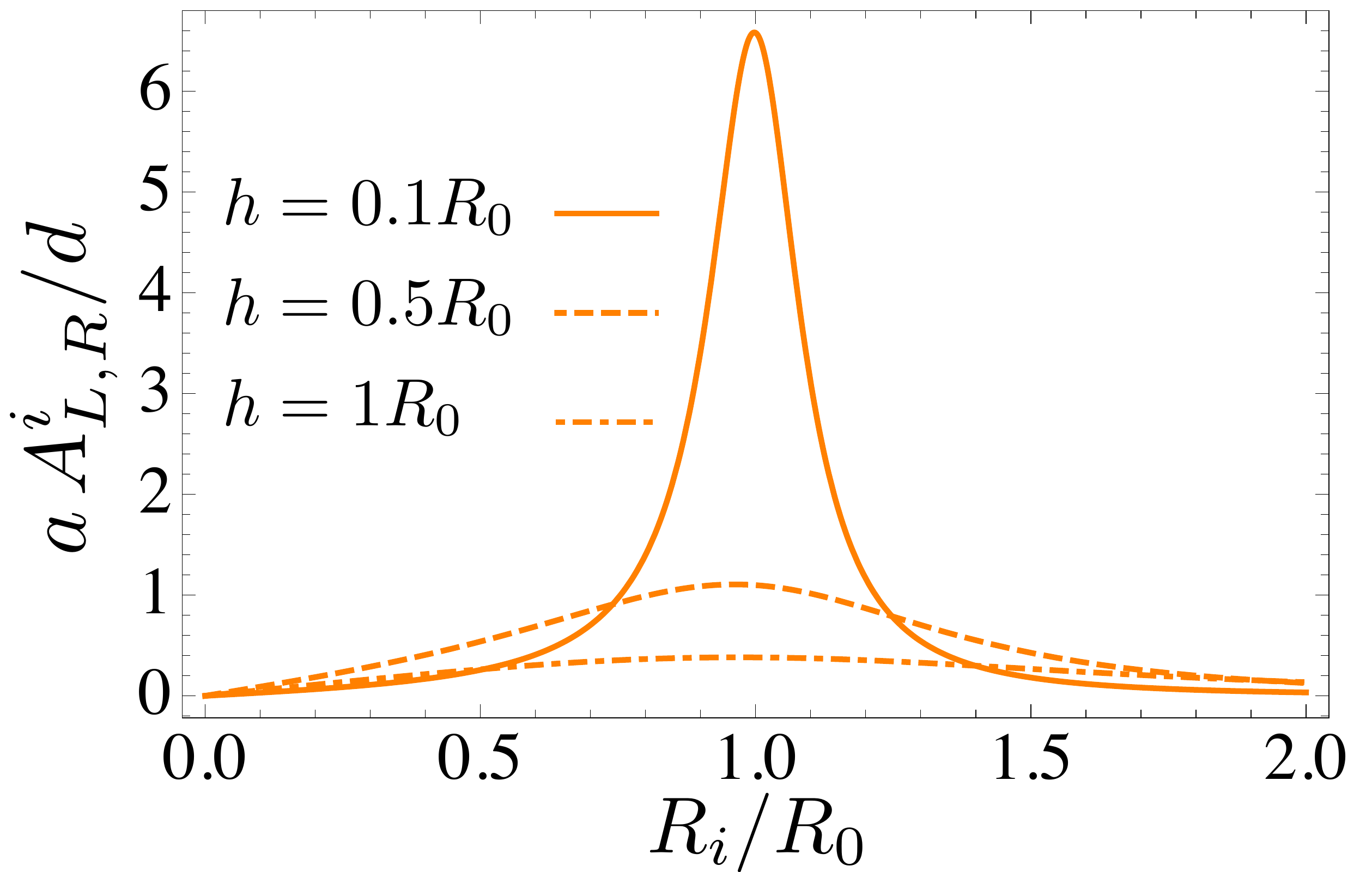}
  \caption{Plot of $aA_{L,R}^i/d$ defined through Eq.~(\ref{eq:constA}) as
  function of $R_i/R_0$ for different values of $h$. We see that the value of
  $aA_{L,R}^i/d$ is bigger when the ST-qubit is closer to the disk of the
  dogbone as expected. Furthermore, by placing the right dot at distance $R_{0}$
  of the disk axis and the left dot on the disk axis, we obtain the strongest
  value for the effective coupling between the two ST-qubits, see Eq.~(\ref{eq:Heff}).}
  \label{fig:Aint}
\end{figure}
Assuming the dogbone shape of the ferromagnet in the above integral and
integration only over the adjacent disk, we obtain
\begin{widetext}
  \begin{align}
    A_{\rm L,R}^i=\frac{2ihd}{a}\frac{2R_{\rm L,R}^iR_0\left( F(\arccsc({w_{\rm L,R}^i}),{w_{\rm L,R}^i}^2)-K({w_{\rm L,R}^i}^2)
    \right)+{u_{\rm L,R}^i} E({w_{\rm L,R}^i}^2)-{u_{\rm L,R}^i} E(\arccsc({w_{\rm L,R}^i}),{w_{\rm L,R}^i}^2)}{3R^i_{\rm L,R}\left( (R_{\rm
    L,R}^i-R_0)^2+h^2\right)\sqrt{(R_{\rm L,R}^i+R_0)^2+h^2}},
    \label{eq:Acalculated}
  \end{align}
\end{widetext}
where $R_0$ is the disk radius, $R_{\rm L,R}^i$ is the distance from the
adjacent disk axis to the left or right quantum well of the $i$-th qubit,
$\arccsc(x)$ is the inverse cosecant; $F(x,y)$, $K(x)$ and $E(x,y)$ are the
corresponding elliptic integrals. Furthermore, we introduced the notation
$u_{\rm L,R}^i={R_{\rm L,R}^i}^2+R_0^2+h^2$ and $w_{\rm
L,R}^i=\sqrt{1-\frac{4R_{\rm L,R}^iR_0}{(R_{\rm L,R}^i+R_0)^2+h^2}}$, where $h$
is the distance in the $z$-direction between the ST-qubit plane and the adjacent
disk bottom and $d$ is the disk thickness, see Fig.~\ref{fig:setup-fig}.

Figure~\ref{fig:Aint} illustrates the dependence of the $A^i_{\rm L,R}$
integrals on the position of the quantum wells. Since the coupling constant is
given by the difference of this integrals for left and right quantum well, we
conclude that the strongest coupling is obtained if one quantum well of the
ST-qubit is positioned below the disk center and the other exactly below the
edge. Furthermore, when $h\ll R_0$ the value of the integral is strongly peaked
around $R\sim R_0$ and this can be exploited as yet another switching
mechanism---moving one quantum well away from the edge of the disk.

\subsection{Sequence for CNOT gate}\label{sec:CNOT}
Two qubits interacting via the ferromagnet evolve according to the Hamiltonian
$H_\mathrm{eff}$, see Eq.~(\ref{eq:Heff}). The Hamiltonian is therefore the sum
of Zeeman terms and qubit-qubit interaction. These terms do not commute, making
it difficult to use the evolution to implement standard entangling gates.
Nevertheless, since $H_\tau$ acts only in the subspace spanned by
$\{\ket{\uparrow\uparrow},\ket{\downarrow\downarrow}\}$ and $\Delta\gg
J_{12}=9\mathcal{B}\chi_\perp(\Delta)/4$, we can neglect the effect of $H_\mathrm{eff}$
in this part of the space and approximate it by its projection in the space
spanned by vectors $\{\ket{\uparrow\downarrow},\ket{\downarrow\uparrow}\}$ 
\begin{equation}
  H'_\mathrm{eff}=H_\tau+J_{12}(\tau_1^x\tau_2^x+\tau_1^y\tau_2^y).
  \label{eq:Heffprime}
\end{equation}
Within this approximation, the coupling in $H'_\mathrm{eff}$ and Zeeman terms
now commute.

We consider the implementation of the iSWAP gate~\cite{imamoglu_quantum_1999}
$U_\mathrm{iSWAP}=e^{-i(\tau_1^x\tau_2^x+\tau_1^y\tau_2^y)3\pi/4}$,
which can be used to implement the CNOT gate:
\begin{equation}\label{eq:iswap}
  U_\mathrm{iSWAP}=e^{iH_\tau t}e^{-iH'_\mathrm{eff}t},
\end{equation}
where $t = 3\pi/(4 J_{12})$. When iSWAP is available, the CNOT gate can be
constructed in the standard way~\cite{tanamoto_efficient_2008}
\begin{equation}\label{eq:cnot}
U_{\rm CNOT}=e^{-i\frac{\pi}{4}\tau_{1}^z}
e^{i\frac{\pi}{4}\tau_{2}^x}
e^{i\frac{\pi}{4}\tau_{2}^z} 
U_{\rm iSWAP} e^{-i\frac{\pi}{4}\tau_{1}^x} U_{\rm iSWAP} 
e^{i\frac{\pi}{4}\tau_{2}^z}.
\end{equation}

Since $H'_\mathrm{eff}$ is an approximation of $H_\mathrm{eff}$, the above
sequence will yield approximate CNOT, $U'_{\rm CNOT}$, when used with the full
Hamiltonian. The success of the sequences therefore depends on the fidelity of
the gates, $F(U'_{\rm CNOT})$. Ideally this would be defined using a
minimization over all possible states of two qubits. However, to characterize
the fidelity of an imperfect CNOT it is sufficient to consider the following
four logical states of two
qubits~\cite{trifunovic_long-distance_2012,2013arXiv1302.4017T}:
$\ket{+,0},\ket{+,1},\ket{-,0},$ and $\ket{-,1}$. These are product states
which, when acted upon by a perfect CNOT, become the four maximally entangled
Bell states $\ket{\Phi^+},\ket{\Psi^+},\ket{\Phi^-},$ and $\ket{\Psi^-}$,
respectively. As such, the fidelity of an imperfect CNOT may be defined,
\begin{equation}
F(U'_{\rm CNOT}) = \min_{i \in \{+,-\}, j \in \{0,1\}} |\bra{i,j} U_{\rm
CNOT}^{\dagger} U'_{\rm CNOT} \ket{i,j}|^2.
\end{equation}
The choice of basis used here ensures that $F(U'_{\rm CNOT})$ gives a good
characterization of the properties of $U'_{\rm CNOT}$ in comparison to a perfect
CNOT, especially for the required task of generating entanglement. For realistic
parameters, with the Zeeman terms two order of magnitude stronger than the
qubit-qubit coupling, the above sequence yields fidelity for the CNOT gate of
$99.976\%$.

To compare these values to the thresholds found in schemes for quantum
computation, we must first note that imperfect CNOT's in these cases are usually
modelled by the perfect implementation of the gate followed by depolarizing
noise at a certain probability. It is known that such noisy CNOT's can be used
for quantum computation in the surface code if the depolarizing probability is
less than $1.1\%$.~\cite{fowler:10} This corresponds to a fidelity, according
to the definition above, of $99.17\%$. The fidelities that may be achieved in
the schemes proposed here are well above this value and hence, though they do
not correspond to the same noise model, we can expect these gates to be equally
suitable for fault-tolerant quantum computation.
\section{Coupling between spin-$1/2$ qubits}\label{sec:onehalf}
In this section we study the coupling of two spin-$1/2$ quantum dots via
interaction with a dog-bone shaped ferromagnet. The Hamiltonian has again the
form as in Eq.~(\ref{eq:Hamiltonian}) and we allow for splittings of the
spin-$1/2$ qubits both along $x$ and $z$ direction,
\begin{equation}\label{eq:Hsigma}
H_{\sigma}=\frac{\Delta_{x}}{2}\sum_{i=1,2}\sigma_{i}^{x}+\frac{\Delta_{z}}{2}\sum_{i=1,2}\sigma_{i}^{z}\,,
\end{equation}
where $\bm{\sigma}_i$ are the Pauli operators of the $i^{\text{th}}$ spin-$1/2$
quantum dot. Hamiltonian (\ref{eq:Hsigma}) is a generalized version of the
Hamiltonian studied in Ref.~\onlinecite{2013arXiv1302.4017T} where we considered
splitting along $x$ only. We present here a detailed derivation of the effective
coupling between two quantum dots located at an arbitrary position with respect
to the dogbone shaped ferromagnet, i.e., contrary to
Ref.~\onlinecite{2013arXiv1302.4017T} we do not assume that the quantum dots are
positioned at a highly symmetric point but consider the most general case. This
allows us to determine the optimal positioning of the qubit in order to achieve
the strongest coupling between the qubits.

The dipolar coupling between the ferromagnet and the spin-$1/2$ qubits is given
by
\begin{equation}
H_{I}=\sum_{i,{\bf r}}\frac{g^{*}\mu_{0}\mu_{B}\mu}{4\pi r^3}\left({\bm \sigma}_{i}\cdot{\bf S}_{\bf r}-\frac{3({\bm \sigma}_{i}\cdot{\bf r})({\bf S}_{\bf r}\cdot{\bf r})}{r^2}\right)\,,
\end{equation}
where $\mu_D$ is the magnetic moment of the spin-$1/2$ qubit. The explicit
expressions for the time evolution of the Pauli operators in Heisenberg picture
is
\begin{eqnarray} \sigma_{i}^{+}(t)&=&-\frac{1}{\Delta^2}(i\Delta\cos(\Delta t/2)-\Delta_z\sin(\Delta t/2))^2\sigma_{i}^{+}\nonumber\\
&&-\frac{\Delta_x^2}{2\Delta^2}(\cos(t\Delta)-1)\sigma_{i}^{-}\nonumber\\
&&+\frac{\Delta_x}{\Delta^2}(\Delta_z-\Delta_z\cos(t\Delta)-i\Delta\sin(t\Delta))\sigma_{i}^z\,,\nonumber\\
\sigma_{i}^z(t)&=&\frac{\Delta_x}{2\Delta^2}(\Delta_z-\Delta_z\cos(t\Delta)-i\Delta\sin(t\Delta))\sigma_{i}^{+}\nonumber\\
&&+\frac{\Delta_x}{2\Delta^2}(\Delta_z-\Delta_z\cos(t\Delta)+i\Delta\sin(t\Delta))\sigma_{i}^-\nonumber\\
&&+\frac{\Delta_z^2+\Delta_x^2\cos(t\Delta)}{\Delta^2}\sigma_{i}^z\,,
\end{eqnarray}
where we introduced the notation $\Delta=\sqrt{\Delta_x^2+\Delta_z^2}$. We also
assume that $\Delta<\Delta_F$ such that the susceptibility
$\chi_\perp(\Delta,\bm r)$ is purely real---thus the transverse noise is gapped.
By replacing the above expressions in Eq.~(\ref{eq:Heffdef}), we obtain the
effective qubit-qubit coupling
\begin{align}
  \label{eq:effegeneral}
H_{\text{eff}}&=H_{\sigma}+\frac{(g^{*}\mu_{0}\mu_{B}\mu)^{2}}{16\pi^2a^6}(\frac{9}{8}A_{1}^{*}A_{2}\chi_{\perp}^{\text{1D}}(\Delta)\sigma_1^z(\Delta)\sigma_{2}^z\nonumber\\
&+\frac{3}{16}(3A_2C_1^{*}\chi_{\perp}^{\text{1D}}(\Delta)-B_1A_2^{*}\chi_{\perp}^{\text{1D}}(\Delta))\sigma_{1}^{+}(\Delta)\sigma_{2}^z\nonumber\\
&+\frac{3}{16}(3A_1C_{2}^{*}\chi_{\perp}^{\text{1D}}(\Delta)-B_2A_1^{*}\chi_{\perp}^{\text{1D}}(\Delta))\sigma_{1}^{z}(\Delta)\sigma_{2}^{+}\nonumber\\
&+\frac{1}{32}(B_1B_2\chi_{\perp}^{\text{1D}}(\Delta)+9C_1C_2^{*}\chi_{\perp}^{\text{1D}}(\Delta))\sigma_{1}^{-}(\Delta)\sigma_{2}^{+}\nonumber\\
&-\frac{3}{32}(B_1C_2\chi_{\perp}^{\text{1D}}(\Delta)+B_2C_1\chi_{\perp}^{\text{1D}}(\Delta))\sigma_{1}^{-}(\Delta)\sigma_{2}^{-}\nonumber\\
&+\text{h.c.})+1\leftrightarrow2\,,
\end{align}
\begin{figure}[h]
  \centering\includegraphics[width=0.9\columnwidth]{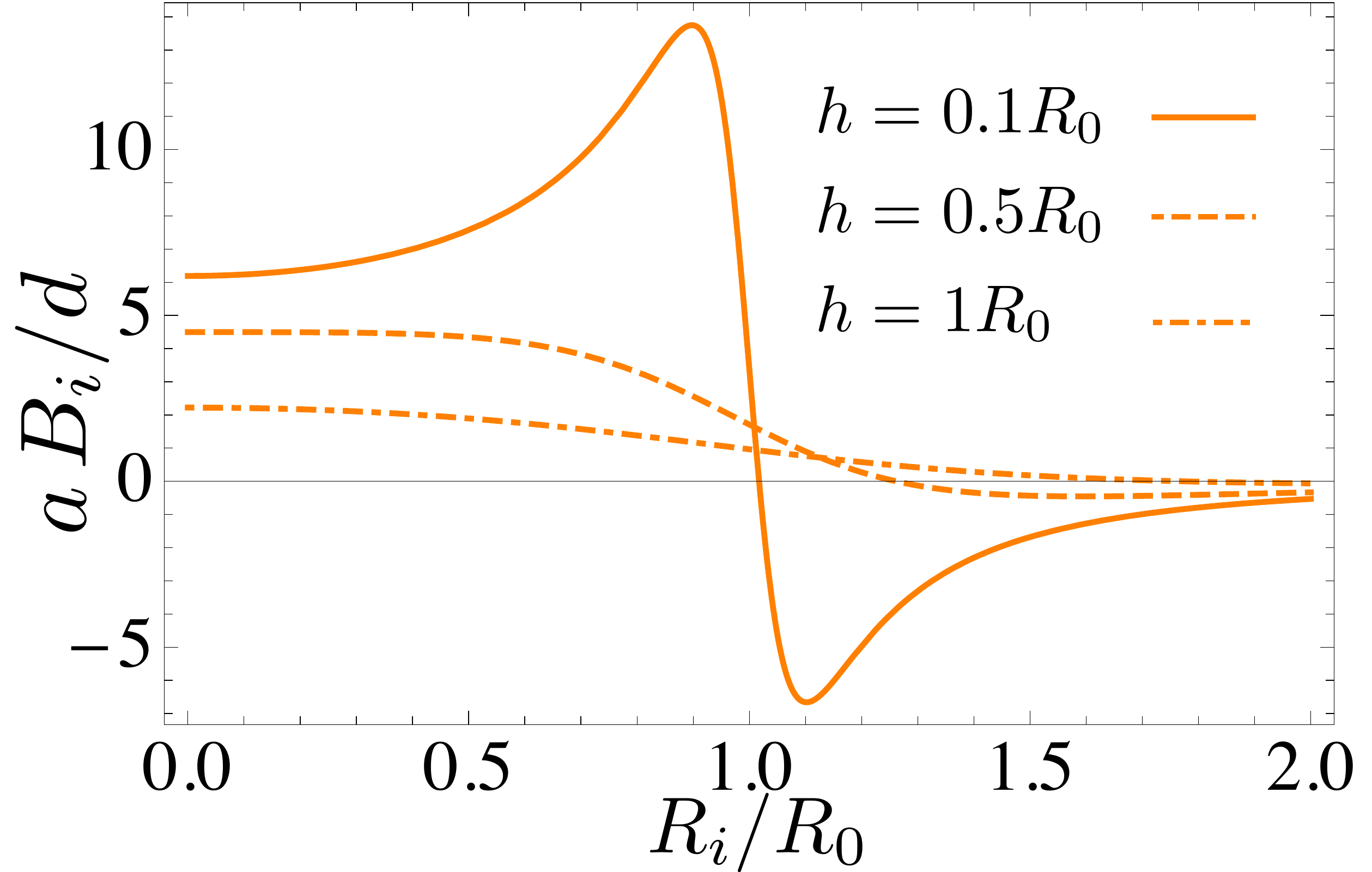}
  \caption{Plot of $aB_{i}/d$ defined in Eq.~(\ref{eq:B}) as function of $R_i/R_0$
  for different values of $h$. The value of the integral increases in general by
  decreasing the value of $h$.}
  \label{fig:Bint}
\end{figure}
\begin{figure}[h]
  \centering\includegraphics[width=0.9\columnwidth]{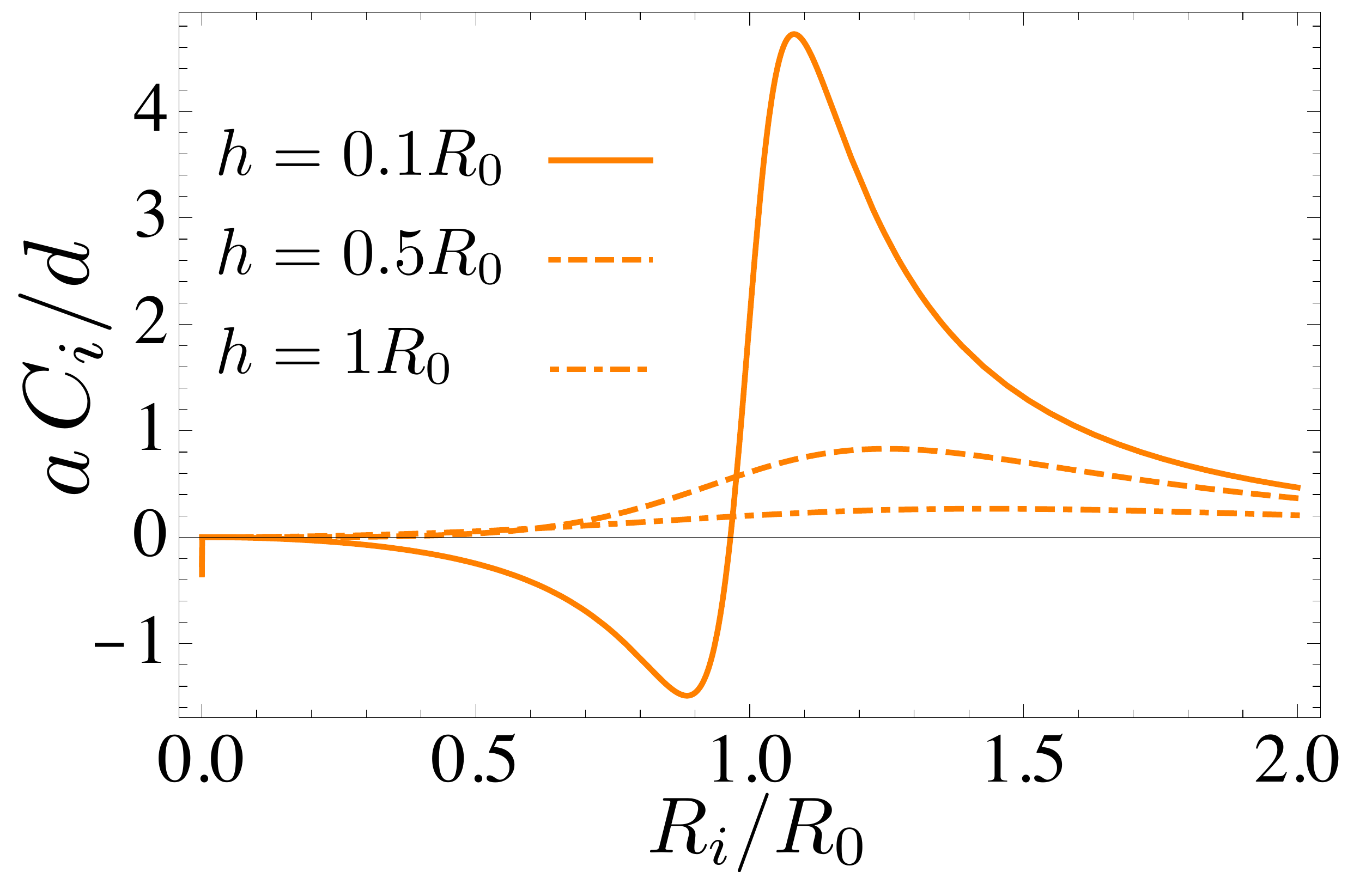}
  \caption{Plot of $aC_{i}/d$ defined in Eq.~(\ref{eq:C}) as function of $R_i/R_0$
  for different values of $h$. The value of the integral is peaked around
  $R_i\sim R_0$ and it increases in general by decreasing the value of
  $h$.}
  \label{fig:Dint}
\end{figure}
where we have denoted $\chi_\perp^{\rm 1D}(\Delta)=\chi_\perp^{\rm
1D}(\Delta,L)$ and introduced the following notation for the integrals
\begin{align}
A_{i}&=\int d{\bf r}_{i}\frac{r^{z}_{i}r^{+}_{i}}{r_{i}^5}\,,\label{eq:A}\\
C_{i}&=\int d{\bf r}_i\frac{(r_{i}^+)^2}{r_i^5}\,,\label{eq:C}\\
B_{i}&=\int d{\bf
r}_{i}\frac{1}{r_{i}^3}\left(2-\frac{3r_{i}^+r_i^-}{r_i^2}\right)\,\label{eq:B},
\end{align}
with the coordinate origin for ${\bf r}_i$ at the $i$-th qubit and the
integration goes over the adjacent disk. We also defined the Fourier transforms
of the time evolution of Pauli matrices $\bm\sigma(t)=e^{iH_\sigma t}\bm\sigma
e^{-iH_\sigma t}$ as
\begin{eqnarray}
\sigma_{i}^{+}(\Delta)&=&-\frac{1}{\Delta^2}\left(-\frac{\Delta^2}{4}+\frac{\Delta_z \Delta}{2}-\frac{\Delta_z^2}{4}\right)\sigma_{i}^+\nonumber\\
&&+\frac{\Delta_x}{\Delta^2}\left(-\frac{\Delta_z}{2}+\frac{\Delta}{2}\right)\sigma_{i}^{z}-\frac{\Delta_x^2}{4\Delta^2}\sigma_{i}^{-}\,,\nonumber
\end{eqnarray}
and
\begin{eqnarray}
\sigma_{i}^z(\Delta)&=&\frac{\Delta_x}{2\Delta^2}\left(-\frac{\Delta_z}{2}+\frac{\Delta}{2}\right)\sigma_{i}^++\frac{\Delta_x}{2\Delta^2}\left(-\frac{\Delta_z}{2}-\frac{\Delta}{2}\right)\sigma_{i}^{-}\nonumber\\
&&+\frac{\Delta_x^2}{2\Delta^2}\sigma_i^z\,.
\end{eqnarray}
By assuming a dogbone-shaped ferromagnet and integrating only over the adjacent
disk as above, we obtain $A_{i}$ given in Eq.~(\ref{eq:Acalculated}) with
$R_{L,R}^{i}$ replaced by $R_i$ since there is now only one spin-$1/2$ qubit
below each disk of the dogbone. The remaining integrals
yield the following results
\begin{widetext}
\begin{eqnarray}
  \label{eq:BCcalculated}
B_i&=&-\frac{2d}{3a}\frac{ \left(R_i^4+3 R_i^2 \left(h^2-R_0^2\right)+ \left(R_0^2+h^2\right)^2\right) E\left(1-w_i^2\right)- \left((R_i-R_0)^2+h^2\right) \left(R_i^2+2 \left(R_0^2+h^2\right)\right) K\left(1-w_i^2\right)}{ R_i^2 \left((R_i-R_0)^2+h^2\right) \sqrt{(R_i+R_0)^2+h^2}}\,,\nonumber\\
C_i&=&\frac{2d}{a}\frac{\left((R_i-R_0)^2+h^2\right) K\left(1-w_i^2\right)-2 \left(R_i^2-R_0^2+h^2\right) E\left(1-w_i^2\right)}{\left((R_i-R_0)^2+h^2\right) \sqrt{(R_i+R_0)^2+h^2}}\,,
\end{eqnarray}
\end{widetext}
where $w_i=\sqrt{1-\frac{4R_i R_0}{(R_i+R_0)^2+h^2}}$, $R_{0}$ is the radius of
each disk, $R_i$ is the distance of the $i$-th qubit to the adjacent dog bone
axis, and $R_0$ and $h$ are defined as in Sec.~\ref{sec:effcoup}. In deriving
Eq.~(\ref{eq:effegeneral}) we took again only `on-resonance' terms into account
(i.e. we neglected $\chi_{\perp}^{\text{1D}}(0)$ and
$\chi_{\perp}^{\text{1D}}(-\Delta)$). Furthermore we assumed, as above, that the
susceptibility between two points on different disks of the dogbone is well
approximated by the 1D transverse susceptibility. In the limit where each
quantum dot lies on the vertical axis going through the center of each cylinder
of the dogbone, the axial symmetry leads to $A_{1}=A_{2}=C_{1}=C_{2}=0$,
$B_{1}=B_{2}=B$, and with $\Delta_z=0$ we recover the result
\begin{equation}
  H_{\text{eff}}=H_{\sigma}+\frac{(g^{*}\mu_{0}\mu_{B}\mu)^{2}}{16\,\pi^2a^{6}}\frac{B^2}{32}\chi^{\rm 1D}_{\perp}(\Delta)(2\sigma_{1}^{y}\sigma_{2}^y+\sigma_{1}^z\sigma_{2}^x+\sigma_{1}^x\sigma_{2}^z)
\end{equation}
derived in Ref.~\onlinecite{2013arXiv1302.4017T}. The analysis carried out
herein assumes arbitrary positioning of the qubit and allow us to determine the
optimal positioning for the strongest coupling. To this end, we analyze
integrals $A_i,\;B_i,\;C_i$, see Figs.~\ref{fig:Aint}-\ref{fig:Dint}. It is
readily observed that the coupling strength increases as the vertical distance
between the qubit and coupler plane, $h$, decreases. Additionally, we observe
that the strongest coupling strength is obtained when the qubit is positioned
below the edge of the adjacent disk.

The derived coupling is valid for any dogbone-like shape of the ferromagnet,
i.e., it is not crucial to assume disk shape.

\subsection{Sequence for CNOT gate}\label{sec:onehalfA}
The effective Hamiltonian derived in previous section,
Eq.~(\ref{eq:effegeneral}), can be re-expressed in the following form
\begin{align}
  H_{\rm eff}=\frac{(g^{*}\mu_{0}\mu_{B}\mu)^{2}}{16\pi^2a^{6}}\chi_\perp^{\rm
  1D}(\Delta,L)\bm\sigma_1^T\cdot\hat{\mathcal
  H}\cdot\bm\sigma_2+\frac{1}{2}\bm\Delta\cdot(\bm
  \sigma_1+\bm\sigma_2),
  \label{eq:Heffgen}
\end{align}
with $\bm\Delta=(\Delta_x,0,\Delta_z)^T$ and $\hat{\mathcal H}$ being the
symmetric matrix with all entries being non-zero. The question now arises how to
construct the CNOT gate sequence for such a general Hamiltonian. We tackle this
problem by taking first the quantization axis to be along the total magnetic
field acting on the two qubits and denote by $\tilde{\bm\sigma}_i$ Pauli matrix
vector with respect to this new quantization axis. The Hamiltonian now reads
\begin{align}
  H_{\rm eff}=\frac{(g^{*}\mu_{0}\mu_{B}\mu)^{2}}{16\pi^{2}a^{6}}\chi_\perp^{\rm
  1D}(\Delta,L)\,\tilde{\bm\sigma}_1^T\cdot\hat{\tilde{\mathcal
  H}}\cdot\tilde{\bm\sigma}_2+\frac{1}{2}\Delta(\tilde{\sigma}_1^z+\tilde{\sigma}_2^z),
  \label{eq:Heffgentilde}
\end{align}
where the components of the matrix $\hat{\tilde{\mathcal H}}$ are given in Appendix \ref{sec:rotated}.

We proceed further along the lines presented in Sec.~\ref{sec:CNOT}, i.e., we
project the rotated Hamiltonian, Eq.~(\ref{eq:Heffgentilde}), on the subspace
spanned by vectors
$\{\ket{\tilde\uparrow\tilde\downarrow},\ket{\tilde\downarrow\tilde\uparrow}\}$.
This procedure yields the following result
\begin{equation}
  H_{\rm eff}^\prime=\tilde
  J_{12}(\tilde{\sigma}_1^x\tilde{\sigma}_2^x+\tilde{\sigma}_1^y\tilde{\sigma}_2^y)+\Delta(\tilde{\sigma}_1^z+\tilde{\sigma}_2^z),
  \label{eq:Heffgenproj}
\end{equation}
$\tilde J_{12}=\frac{(\mu_0 g^{*}\mu_B\mu)^2}{(4\pi)^2a^6}\frac{\chi^{\rm
1D}_\perp(\Delta,L)}{32}\tilde A_{12}$. The dimensionless constant $\tilde
A_{12}$ is defined through the following expression
\begin{align}
  \tilde A_{12}&=\frac{{\Delta_x}^2 \left(36 A_1 A_2+\left(B_1+3 C_1\right)
  \left(B_2+3 C_2\right)\right)}{16 \Delta^2}\nonumber\\
  &+\frac{6 {\Delta_x} {\Delta_z} \left(A_2
  \left(B_1-3 C_1\right)+A_1 \left(B_2-3 C_2\right)\right)}{16
  \Delta^2}\nonumber\\
  &+\frac{2 {\Delta_z}
  \left(B_1 B_2+9 C_1 C_2\right) (\Delta +{\Delta_z})}{16 \Delta^2}.
  \label{eq:Atilde}
\end{align}

The projected Hamiltonian in Eq.~(\ref{eq:Heffgenproj}) is identical to the one
already considered in Sec.~\ref{sec:CNOT}, Eq.~(\ref{eq:Heffprime}). Thus the
CNOT gate sequence can be obtained in exactly same way, namely via
Eqs.~(\ref{eq:iswap}) and (\ref{eq:cnot}).

Similar to the previously studied case of ST-qubits, the CNOT gate sequence
described in this section is only approximate one. For realistic parameters,
with the Zeeman terms two order of magnitude stronger than the qubit-qubit
coupling, this approximate sequence yields fidelity for the CNOT gate similar to
the one previously found in Sec.~\ref{sec:CNOT}.

\begin{figure}[h]
  \centering\includegraphics[width=0.9\columnwidth]{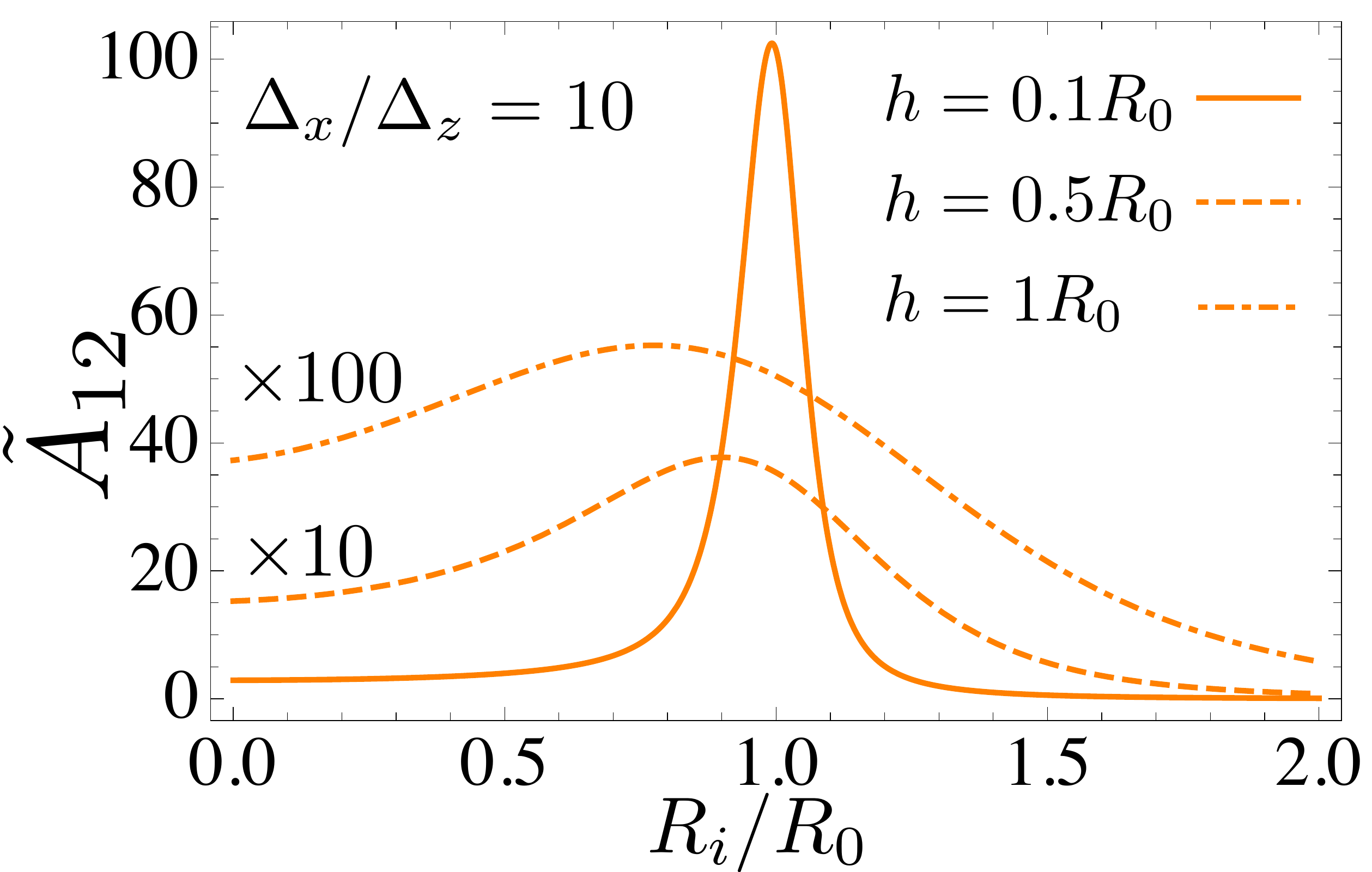}
  \caption{Plot of $\tilde A_{12}$ defined in Eq.~(\ref{eq:Atilde}) as function
  of $R_i/R_0$ for different values of $h$, assuming $R_1=R_2$ and
  $\Delta_x=10\Delta_z$. The value of the integral is peaked around $R_i\sim
  R_0$ and it increases in general by decreasing the value of $h$.}
  \label{fig:A12big}
\end{figure}

\begin{figure}[h]
  \centering\includegraphics[width=0.9\columnwidth]{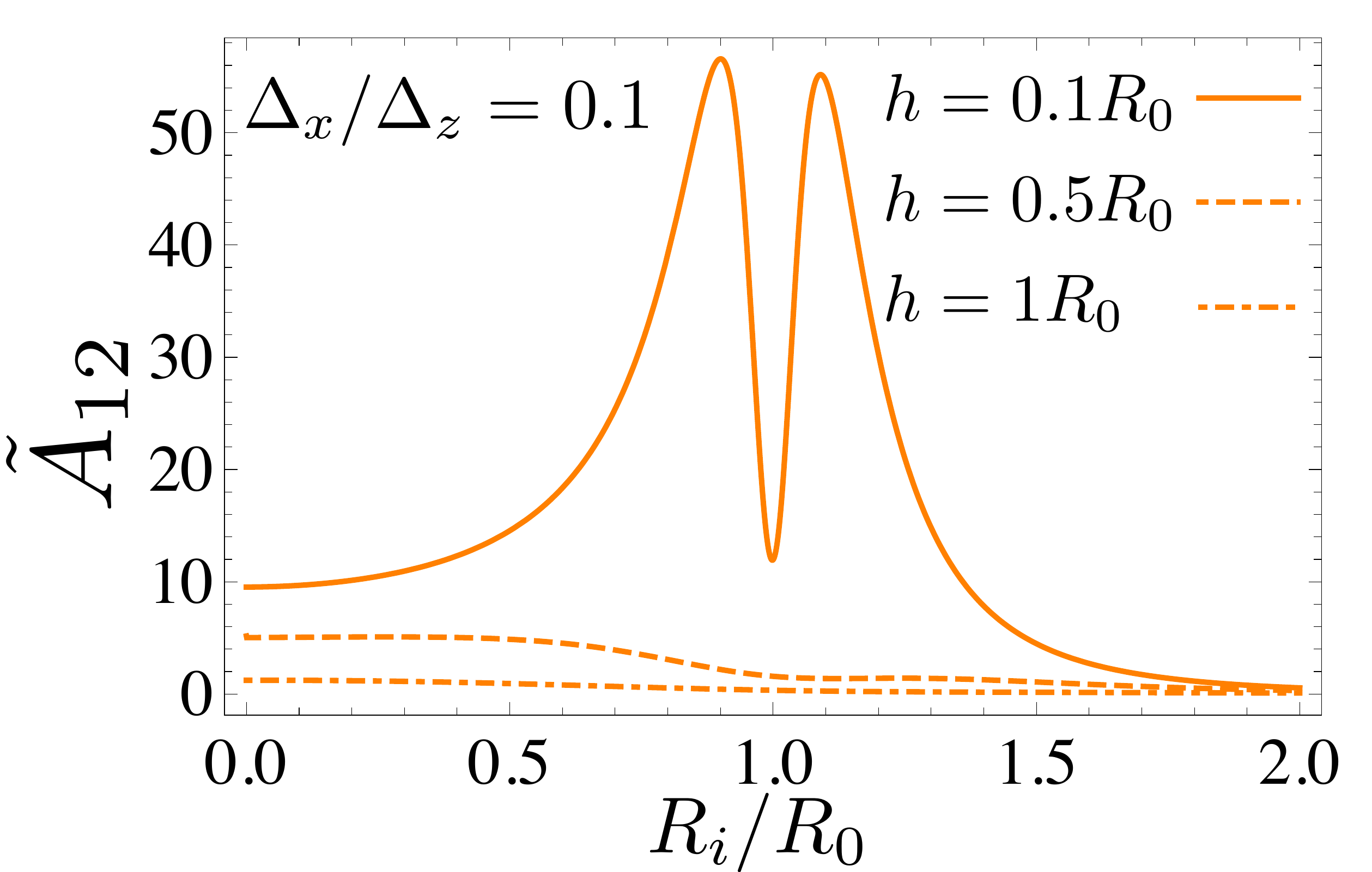}
  \caption{Plot of $\tilde A_{12}$ defined in Eq.~(\ref{eq:Atilde}) as function
  of $R_i/R_0$ for different values of $h$, assuming $R_1=R_2$ and
  $\Delta_x=0.1\Delta_z$. The value of the integral is peaked around $R_i\sim
  R_0$ only for $h\ll R_0$ and it increases in general by decreasing the value
  of $h$.}
  \label{fig:A12small}
\end{figure}

We now use Eq.~(\ref{eq:Atilde}) to determine the optimal positioning of the
qubits in order to obtain shortest possible gate operation times. If we assume
that the qubit splitting is predominantly along the $x$-axis
($\Delta_x\gg\Delta_z$), we obtain the behavior illustrated in
Fig.~\ref{fig:A12big}. We conclude that for all values of $h$ the optimal
positioning is below the edge of the adjacent disk. It is interesting to note
that when $h\ll R_0$ one can obtain more than two orders of magnitude
enhancement compared to the positioning previously studied in
Ref.~\onlinecite{2013arXiv1302.4017T}. In the opposite limit,
$\Delta_x\ll\Delta_z$, we observe behavior illustrated in
Fig.~\ref{fig:A12small}. When also $h\ll R_0$ we recover the same optimal
positioning as before---below the edge of the disk, while when $h\sim R_0$,
positioning the qubit anywhere below the disk yields approximately same coupling
strength.

\section{Coupling between spin-$1/2$ and ST-qubits}\label{sec:Hybrid}
In the previous sections we have considered the coupling of both spin-$1/2$ and
ST qubits individually. Since each setup has its own advantages and
challenges, it is interesting to show these qubits can be cross-coupled to each
other and thus that hybrid spin-qubits can be formed. This opens up the
possibility to take advantage of the ’best of both worlds’.

The Hamiltonian of such a hybrid system reads
\begin{equation}
  H=H_F+H_\sigma+H_\tau+H_I,
  \label{eq:Hhybrid}
\end{equation}
where the first three term on left-hand side are given by omitting the summation
over $i$ in Eqns.~(\ref{eq:ferromagnet}),(\ref{eq:Hsigma}) and (\ref{eq:qubitH}),
respectively. The interaction term $H_I$ has the following form
\begin{align}
H_I&=\sum_{{\bf r}}\frac{g^{*}\mu_{0}\mu_{B}\mu}{4\pi r^3}\left({\bm \sigma}\cdot{\bf
S}_{\bf r}-\frac{3({\bm \sigma}\cdot{\bf r})({\bf S}_{\bf r}\cdot{\bf
r})}{r^2}\right)\nonumber\\
&+g^{*}\mu_{B}\left( \hat B_{\rm L}^z-\hat B_{\rm R}^z \right)\tau_i^x,
\label{eq:HIhybrid}
\end{align}
with $\hat B_{\rm L,R}$ being given in Eq.~(\ref{eq:dipolarB}) when index
$i$ is omitted. Continuing along the lines of the previous sections, we perform
the second order SW transformation and obtain the effective coupling between the
qubits

\begin{align}
  H_{\rm
  eff}&=\frac{3(\mu_{0}g^{*}\mu_{B}\,\mu)^{2}}{256\pi^{2} a^{6}}\chi_\perp^{\text{1D}}(\Delta)\left(
  \mathrm{Re}[3{A(A_L-A_R)^*}]\tau^x\sigma^z(\Delta)\right.\nonumber\\
  &+\left.(C^*(A_L-A_R)-3B(A_L-A_R)^*)\tau^x\sigma^+(\Delta)+\text{h.c}\right)\nonumber\\
  &+\{\sigma^i(\Delta)\rightarrow\sigma^i,\tau^x\rightarrow\tau^x(\Delta)\},
  \label{eq:Heff_hybrid}
\end{align}
where $A_{\rm L,R}$ and $A$ are calculated in Eq.~(\ref{eq:Acalculated}), while
$B$ and $C$ are given in Eq.~(\ref{eq:BCcalculated}).

Similarly as in the previous sections, we find that the optimal coupling for the
hybrid case is obtained when the spin-$1/2$ qubit is positioned below the edge of one of the
two discs while one quantum well of the ST-qubit is positioned below the other
disc center with the other well being below the disc edge.

\section{Validity of the effective Hamiltonian}\label{sec:validity}
We discuss herein the validity of the effective Hamiltonian derived in
Sec.~\ref{sec:STC} and Sec.~\ref{sec:onehalf}. In deriving those effective
Hamiltonians we assumed that the state of the ferromagnet adapts practically
instantaneously to the state of the two qubits at a given moment. This is true
only if the dynamics of the two qubits is slow enough compared to the time scale
for the ferromagnet to reach its equilibrium. To estimate the equilibration time
of the ferromagnet we use the phenomenological Landau-Lifshitz-Gilbert
equation~\cite{tserkovnyak_nonlocal_2005} in standard notation
\begin{equation}
  \partial_t\bm m=-\gamma\bm m\times\bm H_{\rm eff}+\alpha\bm m\times\partial_t\bm m,
  \label{eq:LLG}
\end{equation}
where $\bm m(\bm r,t)$ is the magnetization of the (classical) ferromagnet at
the given time $t$ and spatial coordinate $\bm r$; $\gamma=2\mu_{\rm B}/\hbar$
is the gyromagnetic ratio and $\alpha$ is Gilbert damping constant. The above
equation has to be supplemented by an equation for the effective magnetic field,
$\bm H_{\rm eff}(\bm r,t)$.

However, here we do not aim at studying the exact dynamics of the ferromagnet
but rather at giving a rough estimate of its equilibration time scale. We first estimate
the time needed for the effective field, $\bm H_{\rm eff}(\bf r,t)$, to reach
its final orientation along which the magnetization of the ferromagnet, $\bm m(\bf r,t)$, will
eventually point. This time is roughly given by the ratio of the ferromagnet size to the
relevant magnon velocity. The typical magnon energy is given by $A$, and thus
the relevant magnon velocity can be estimated as  $v_M=\sqrt{J S A}\,a/\hbar$, leading to
times of $10ns$ for magnons to travel over distances of $1\mu m$ in
ferromagnets. After $\bm H_{\rm eff}(\bf r,t)$ has reached its final value, it
takes additional time for $\bm m(\bf r,t)$ to align along it. This
time can be estimated from Eq.~(\ref{eq:LLG}), as $T_p/\alpha$, where $T_p$ is
the precession time around the effective magnetic field which can be
approximated by the externally applied magnetic field (leading to the gap
$\Delta_F$). Using the typical Gilbert damping factor of
$\alpha=0.001$~\cite{tserkovnyak_nonlocal_2005} we arrive at the time scale of
$10ps$. Therefore, the total time for the ferromagnet considered herein to reach
its equilibrium is about $10ns$. This in turn leads to the bottleneck for the
operation time.

Special care has to be taken for the validity of the perturbation theory
employed herein, since we are working close to resonance, i.e.,
$\Delta-\Delta_F$ has to be small but still much larger than the coupling of a
qubit to an individual spin of the ferromagnet. For the perturbation theory to
be valid we also require the tilt of each ferromagnet spin  to be sufficiently small (i.e.
$\langle S_{\bf r}^\pm\rangle\ll 1$). The tilt of the central spin of the
ferromagnetic disk can be estimated by the integral over the dogbone disk $D$
\begin{equation}
\langle S^{\pm}_{\bf r}\rangle=\int_{D}\chi_{\perp}({\bf r}) B_{\perp}(\bf r)\,.
\end{equation}
Using cylindrical coordinates we then obtain
\begin{equation}\label{eq:tilting}
  \langle S_{\bf r}^\pm\rangle\sim\frac{\mu_0\mu_B^2}{2
  a}\int_0^R\rho d\rho\frac{1}{(\rho^2+h^2)^{3/2}}\frac{S}{D\rho},
\end{equation}
where $\frac{S}{D\rho}$ is the spatial decay of the transversal susceptibility
and $\frac{1}{(\rho^2+h^2)^{3/2}}$ is the decay of the dipolar field causing the
perturbation of the ferromagnet. Even though each spin is just slightly tilted, we
obtain a sizable coupling due to big number of spins involved in mediating the
coupling.

\section{Switching mechanisms}\label{sec:switching} In this
section we briefly discuss possible switching \textit{on/off} mechanisms. These
include changing the splitting of the qubits and moving them spatially. The
former mechanism is based on the dependence of the susceptibility decay length
on frequency,~\cite{2013arXiv1302.4017T} see Eq.~(\ref{eq:chi_1D_r}). It is
enough to detune the qubit splitting by less then a percent to switch the
qubit-qubit coupling effectively off. This is particularly feasible for the
ST-qubits where qubit splitting can be controlled by all electrical means.
Furthermore, the ST-qubits coupling can be switched off also by rotating them
such that $A_{\rm L}=A_{\rm R}$, see Eq.~(\ref{eq:constA}).

The spin-$1/2$ qubits can be switched either by detuning its splitting
off-resonance with the magnon gap $\Delta_F$ or by moving them away from the
dogbone disk, see Figs.~\ref{fig:A12big}-\ref{fig:A12small}.

\section{Coupling Strengths and Operation times}\label{sec:Operation_Times}
In Tables \ref{tab:1} and \ref{tab:2} we present a summary of the effective
coupling strengths and operation times that can be obtained in the proposed
setup. We assume that the qubits are separated by a distance of $1\mu m$ and we
give the remaining parameters in the table captions. 

The column captions correspond to four experimentally relevant setups
considered in this work (GaAs ST and spin-$1/2$ quantum dots, silicon-based
quantum dots, and NV-centers). The row captions denote respectively the
vertical distance $h$ between the qubit and the disk of the ferromagnet, the
difference between the qubit splitting $\Delta$ and the internal splitting
$\Delta_{F}$ of the ferromagnet (given in units of energy and in units of
magnetic field), the obtained effective qubit-qubit interaction, and the
corresponding operation time.

The operation times obtained in Tables \ref{tab:1} and \ref{tab:2} are
significantly below the relaxation and decoherence times of the corresponding
qubits. Indeed, for GaAs quantum dots $T_1=1s$ (see
Ref.~\onlinecite{amasha_electrical_2008}), and $T_2>200\mu s$ (see
Ref.~\onlinecite{Bluhm_Dephasing_2011}), respectively.  Here we compare to
$T_2$ instead of $T_2^*$ since spin-echo can be performed together with
two-qubit gates.~\cite{khodjasteh_dynamically_2009} Alternatively, the $T_2^*$
of GaAs qubits can be increased without spin-echo by narrowing the state of the
nuclear spins.~\cite{Xu_Optically_2009,Vink_Locking_2009}

For silicon-based qubits decoherence time up to $T_{2}\approx200\mu s$ is
achievable.~\cite{pla_high_2013} Finally decoherence times of $T_{2}^{*}\approx
20\mu s$ and $T_{2}\approx 1.8 ms$ have been obtained for N-V centers in
diamond.~\cite{balasubramanian2009ultralongspin}

In Table \ref{tab:12}, we summarize the obtained coupling strengths and
operation times obtained when a ST-qubit is cross coupled with a spin-$1/2$
qubit.

We have verified that the tilting of the ferromagnet spins given in
Eq.~(\ref{eq:tilting}) remains small. The biggest tilt we obtain (for $h=5\,nm$) is
$\langle S_{\bf r}^{\pm}\rangle\approx 10^{-7}\ll1$. Thus all the result are
within the range of validity of the perturbation theory.
\begin{widetext}
\begin{center}
\begin{table}
     \caption{The parameters used to obtain the numbers below are: Land\'e factor of the ferromagnet $g_{F}=2$;  disk radius $R_{0}=50nm$;  disk thickness $d=20nm$ ; Curie temperature $T=550K$ and thus exchange coupling $J/k_{B}\approx 824 K$; lattice constant of the ferromagent $a=4\angstrom$. We consider the case $\Delta_{x}\ll\Delta_{z}$.}
     \label{tab:1}
\begin{tabular}{||c|ccccc||} 
   \hline
   \hline
  &\multicolumn{1}{c|}{ GaAs ST QD} &\multicolumn{1}{c|} {GaAs ST QD} &\multicolumn{1}{c|} {GaAs spin-$1/2$ QD} &\multicolumn{1}{c|}{Silicon-based QD} & NV-center \\
    $\Delta_{x}\ll\Delta_{z}$&\multicolumn{1}{c|}{ $\vert g^{*}\vert=0.4$} & \multicolumn{1}{c|}{$\vert g^{*}\vert=0.4$} &\multicolumn{1}{c|} {$\vert g^{*}\vert=0.4$} &\multicolumn{1}{c|}{ $\vert g^{*}\vert=2$} & $\vert g^{*}\vert=2$\\
    \hline
    \hline
   Distance $h$ & \multicolumn{1}{c|}{$50\, nm$} &\multicolumn{1}{c|}{ $50\, nm$} &\multicolumn{1}{c|}{ $50\, nm$} &\multicolumn{1}{c|}{ $25\,nm$} & $5\,nm$\\
   \hline
   Splitting $ \Delta_{F}-\Delta$ &\multicolumn{1}{c|}{ $1\,\mu eV$ ($43.2\, mT$)} &\multicolumn{1}{c|}{ $0.5 \mu eV$ ($21.6 mT$)} &\multicolumn{1}{c|}{ $10^{-2} \,\mu eV$ ($0.4\, mT$)} &\multicolumn{1}{c|}{ $10^{-2} \,\mu eV$ ($0.1\,mT$)} & $10^{-1}\,\mu eV$ ($0.9\,mT$) \\
   \hline
  Coupling strength (CS) &\multicolumn{1}{c|}{ $1.4\times10^{-9} \,eV$} &\multicolumn{1}{c|}{$1.4\times 10^{-8}$} &\multicolumn{1}{c|}{ $2\times 10^{-10} \,eV$} &\multicolumn{1}{c|}{ $2.4\times 10^{-8}\, eV$} & $1.8\times 10^{-8}\,eV$\\
  \hline
  Operation time (OS)& \multicolumn{1}{c|}{$470 \,ns$} &\multicolumn{1}{c|}{ $47 ns$} &\multicolumn{1}{c|}{$3.3\, \mu s$} &\multicolumn{1}{c|}{$ 27.4\, ns$} & $36.6\, ns$\\
  \hline
  \hline
    \end{tabular}
    \end{table}
\begin{table}
     \caption{We use the same parameters as in Table~\ref{tab:1} but consider the case $\Delta_{x}\gg\Delta_{z}$.}
     \label{tab:2}
\begin{tabular}{||c|ccc||} 
\hline
\hline
& \multicolumn{1}{c|}{GaAs spin-$1/2$ QD} & \multicolumn{1}{c|}{Silicon-based QD} & NV-center \\
$\Delta_{x}\gg\Delta_{z}$ & \multicolumn{1}{c|}{$\vert g^{*}\vert=0.4$} &\multicolumn{1}{c|}{ $\vert g^{*}\vert=2$} & $\vert g^{*}\vert=2$\\
\hline
\hline
Distance $h$ &\multicolumn{1}{c|}{ $50\, nm$} & \multicolumn{1}{c|}{$25\,nm$} & $5\,nm$\\
\hline
Splitting $ \Delta_{F}-\Delta$ &\multicolumn{1}{c|}{ $10^{-2} \,\mu eV$ ($0.4\, mT$)} &\multicolumn{1}{c|}{ $10^{-2} \,\mu eV$ ($0.1\,mT$)} & $10^{-1}\,\mu eV$ ($0.9\,mT$) \\
\hline
Coupling strength  &\multicolumn{1}{c|}{ $1.2\times 10^{-10} \,eV$} & \multicolumn{1}{c|}{$1.8\times 10^{-8}\, eV$} & $3.6\times 10^{-8}\,eV$\\
\hline
Operation time &\multicolumn{1}{c|}{  $5.5\, \mu s$} &\multicolumn{1}{c|}{$ 36.6\, ns$} & $18.3\, ns$\\
\hline
\hline
\end{tabular}
\end{table}
\begin{table}\label{tab:12}
  \caption{We use the same parameters as in Table~\ref{tab:1} and choose the
  splitting $\Delta_{F}-\Delta=10^{-2}\,\mu eV$ for the ST-qubit (the splitting
  of the other qubit is taken from Table~\ref{tab:1}) to determine the
  coupling strengths and operation times achieved in the hybrid case. The
  column caption of the table labels GaAs ST-QD, while the row captions label
  the three other qubit systems, considered in this work, to which it can be
  hybridized. The left panel corresponds to the case $\Delta_{x}\ll\Delta_{z}$
  while the right panel corresponds to $\Delta_{x}\gg\Delta_{z}$.}
  \begin{tabular}{||c|cc||}
  \hline
  \hline
  $\Delta_{x}\ll\Delta_{z}$  &\multicolumn{2}{c||}{ GaAs ST QD}  \\
  \hline
  &\multicolumn{1}{c|}{ Coupling strength} & Operation time\\
  \hline
  GaAs spin-$1/2$ QD &\multicolumn{1}{c|}{$1.7\times 10^{-9}\,eV$ } & $387\,ns$ \\
  \hline
  Silicon-based QD &\multicolumn{1}{c|}{$1.8\times 10^{-8}\,eV$ }  & $36.6\,ns$ \\
  \hline
  NV-center &\multicolumn{1}{c|}{$1.6\times 10^{-8}\,eV$ }  & $41.1\,ns$ \\
  \hline
  \hline
  \end{tabular}
  \begin{tabular}{||c|cc||}
  \hline
  \hline
  $\Delta_{x}\gg\Delta_{z}$  &\multicolumn{2}{c||}{ GaAs ST QD}  \\
  \hline
  &\multicolumn{1}{c|}{ Coupling strength} & Operation time\\
  \hline
  GaAs spin-$1/2$ QD &\multicolumn{1}{c|}{$1.3\times 10^{-9}\,eV$ } & $506\,ns$ \\
  \hline
  Silicon-based QD &\multicolumn{1}{c|}{$1.6\times 10^{-8}\,eV$ }  & $41.1\,ns$ \\
  \hline
  NV-center &\multicolumn{1}{c|}{$2.2\times 10^{-8}\,eV$ }  & $29.9\,ns\,ns$ \\
  \hline
  \hline
  \end{tabular}
\end{table}

\end{center}
\end{widetext}

\section{Conclusions}\label{sec:conclusions}
We have proposed and studied a model that allows coherent coupling of distant
spin qubits. The idea is to introduce a piece of ferromagnetic material between
qubits to which they couple dipolarly. A dogbone shape of the ferromagnet is
the best compromise since it allows both strong coupling of the qubits to the
ferromagnet and long-distance coupling because of its slowly decaying 1D
spin-spin susceptibility. We have derived an effective Hamiltonian for the
qubits in the most general case where the qubits are positioned arbitrarily
with respect to the dogbone. We have calculated the optimal position for the
effective qubit-qubit coupling to be strongest and estimated it. For both the
singlet-triplet (ST) and spin-$1/2$ qubits, interaction strengths of $10^{-2}{\rm \mu
eV}$ can be achieved. Since decoherence effects induced by the coupling to the
ferromagnet are negligible,~\cite{2013arXiv1302.4017T} we obtain error
thresholds of about $10^{-4}$ for ST-qubits and for spin-$1/2$
qubits. In both cases this is good enough to implement the surface code error
correction.~\cite{raussendorf_fault-tolerant_2007} Finally, for both types of
qubits we have explicitly constructed the sequence to implement a CNOT gate
achievable with a fidelity of more than 99.9\%

Our analysis is general and is not restricted to any special types of qubits as
long as they couple dipolarly to the ferromagnet. Furthermore, the only relevant
quantity of the coupler is its spin-spin susceptibility. Hence, our analysis is
valid for any kind of coupler (and not just a ferromagnet) that has a
sufficiently slowly decaying susceptibility. 

This quantum computing architecture retains all the single qubit gates and
measurement aspects of earlier approaches, but allows qubit spacing at
distances of order 1 $\mu m$ for two- qubit gates, achievable with the
state-of-the-art semiconductor technology.

\section{Acknowledgment}
We would like to thank A.~Yacoby, A.~Morello, and C.~Kloeffel for useful discussions. This
work was supported by the Swiss NSF, NCCR QSIT, and IARPA.
\appendix
\section{Holstein-Primakoff transformation}
For the sake of completeness we derive in this Appendix explicit expressions for
the different spin-spin correlators used in this work
\begin{equation}
C^{\alpha\beta}(\omega,{\bf q})=\langle S_{{\bf q}}^{\alpha}(\omega)S_{-{\bf q}}^{\beta}(0)\rangle\,.
\end{equation}

For this purpose, we make use of a Holstein-Primakoff
transformation
\begin{eqnarray}
S_{i}^{z}&=&-S+n_{i},\,\,\,S_{i}^{-}=\sqrt{2S}\sqrt{1-\frac{n_i}{2S}}a_i,\,\,\,\mathrm{and}\nonumber\\
S_{i}^{+}&=&\left(S_{i}^{-}\right)^{\dagger},
\end{eqnarray}
in the limit $n_{i}\ll 2S$, with $a_{i}$ satisfying bosonic commutation
relations and $n_{i}=a_{i}^{\dagger}a_{i}$ \cite{nolting2009quantumtheory}. The
creation operators $a_{i}^{\dagger}$ and annihilation operators $a_{i}$ satisfy
bosonic commutation relations and the associated particles are called magnons.
The corresponding Fourier transforms are straightforwardly defined as $a_{{\bf
q}}^{\dagger}=\frac{1}{\sqrt{N}}\sum_{i}e^{-i{\bf q}\cdot{\bf R}_{i}}a_{i}$. In
harmonic approximation, the Heisenberg Hamiltonian $H_{F}$ reads
\begin{equation}
H_{F}\approx\sum_{\bf q}\epsilon_{{\bf q}}a_{{\bf q}}^{\dagger}a_{{\bf q}}\,,
\end{equation}
where $\epsilon_{\bf q}=\omega_{\bf
q}+\Delta_F=4JS[3-(\cos(q_x)+\cos(q_y)+\cos(q_z))]+\Delta_{F}$ is the spectrum
for a cubic lattice with lattice constant $a=1$ and the gap $\Delta_{F}$ is
induced by the external magnetic field or anisotropy of the ferromagnet.

\section{Transverse correlators $\langle S_{{\bf q}}^{+}(t)S_{-{\bf q}}^{-}(0)\rangle$}
Let us now define the Fourier transforms in the harmonic approximation
\begin{eqnarray}
S_{{\bf q}}^{+}&=&\frac{1}{\sqrt{N}}\sum_{i}e^{-i{\bf q}\cdot{\bf r}_{i}}S_{i}^{+}\approx \frac{\sqrt{2S}}{\sqrt{N}}\sum_{i}e^{-i{\bf q}\cdot{\bf r}_{i}}a_i^{\dagger}=\sqrt{2S}a_{-{\bf q}}^{\dagger}\nonumber\,,\\
S_{-{\bf q}}^{-}&=&\frac{1}{\sqrt{N}}\sum_{i}e^{i{\bf q}\cdot{\bf r}_{i}}S_{i}^{-}\approx \frac{\sqrt{2S}}{\sqrt{N}}\sum_{i}e^{i{\bf q}\cdot{\bf r}_{i}}a_i=\sqrt{2S}a_{-{\bf q}}\,.
\end{eqnarray}
From this it directly follows that
\begin{align}
C^{+-}(t,{\bf q})&=\langle S_{{\bf q}}^{+}(t)S_{-{\bf
q}}^{-}(0)\rangle\nonumber\\
&=2S\langle
a_{-{\bf q}}^{\dagger}(t)a_{-{\bf q}}\rangle=2S e^{i\epsilon_{\bf q}t}n_{{\bf
q}}\,,
\end{align}
with $\epsilon_{\bf q}\approx D{\bf q}^2+\Delta_F$ in the long-wavelength
approximation.

The Fourier transform is then given by
\begin{eqnarray}
C^{+-}(\omega,{\bf q})&=&\frac{1}{\sqrt{2\pi}}\int_{-\infty}^{\infty}dt
e^{-i\omega t}C^{+-}(t,{\bf q})\nonumber\\
&=&\underbrace{\frac{1}{\sqrt{2\pi}}\int_{-\infty}^{\infty}dte^{i(\epsilon_{\bf
q}-\omega)t}}_{\sqrt{2\pi}\delta(\epsilon_{\bf q}-\omega)}2S n_{\bf
q}\nonumber\\
&=&\sqrt{2\pi}2S\delta(\epsilon_{\bf q}-\omega)\frac{1}{e^{\beta\omega}-1}\,.
\end{eqnarray}
The corresponding correlator in real space becomes 
($q=\vert{\bf q}\vert$)
\begin{eqnarray}
C^{+-}(\omega,{\bf r})&=&\frac{1}{(2\pi)^{3/2}}\int d{\bf q}e^{i{\bf q}\cdot{\bf
r}}C^{+-}(\omega,{\bf q})\\
&=&\frac{\sqrt{2\pi}}{(2\pi)^{3/2}}2S\frac{1}{e^{\beta\omega}-1}\int d{\bf q}\delta(D{\bf q}^2+\Delta_F-\omega)e^{i{\bf q}\cdot{\bf r}}\nonumber\\
&=&\frac{2S}{e^{\beta\omega}-1}\int_{-1}^{1}\int_{0}^{\infty}dqdxq^2\delta(D q^2+\Delta_F-\omega)e^{iq rx}\nonumber\\
&=&\frac{4S}{r}\frac{1}{e^{\beta\omega}-1}\int_{0}^{\infty}dq q \delta(Dq^2+\Delta_F-\omega)\sin(qr)\,.\nonumber
\end{eqnarray}
Let us now perform the following substitution
\begin{equation}
y=Dq^2,
\end{equation}
which gives for $\omega>\Delta_F$
\begin{align}
C^{+-}(\omega,{\bf r})&=\frac{4S/r}{2D(e^{\beta\omega}-1)}\int_{0}^{\infty}dy\delta(y+\Delta_F-\omega)\times\nonumber\\
&\times\sin\left(\sqrt{\frac{y}{D}}r\right)\\
&=\frac{2S}{D}\frac{1}{e^{\beta\omega}-1}\frac{\sin(\sqrt{(\omega-\Delta_F)/D}r)}{r}\,\nonumber.
\end{align}
We remark  that
\begin{equation}
C^{+-}(\omega,{\bf r})=0,\quad\omega<\Delta_F.
\end{equation}
We note the diverging behavior of the above correlation function for $\Delta_F=0$
and $\omega\rightarrow0$, namely
\begin{eqnarray}
\lim\limits_{\omega\rightarrow0}\frac{1}{e^{\beta\omega}-1}\frac{\sin\left(\sqrt{\frac{\omega}{D}r}\right)}{r}&=&\lim\limits_{\omega\rightarrow0}\frac{\sqrt{\omega/D}r}{r\beta\omega}\nonumber\\
&&\rightarrow \frac{1}{\sqrt{D}\beta}\frac{1}{\sqrt{\omega}}\,.
\end{eqnarray}
Similarly, it is now easy to calculate the corresponding commutators and
anticommutators. Let us define
\begin{eqnarray}
S_{\perp}(t,{\bf q}):=\frac{1}{2}\{S_{{\bf q}}^{+}(t),S_{-{\bf q}}^{-}(0)\}\,.
\end{eqnarray}
It is then straightforward to show that
\begin{eqnarray}
S_{\perp}(t,{\bf q})&=&Se^{i\epsilon_{\bf q}t}(1+2n_{{\bf q}})\,,
\end{eqnarray}
and therefore
\begin{eqnarray}
S_{\perp}(\omega,{\bf q})&=&\frac{S}{\sqrt{2\pi}}\int_{-\infty}^{\infty}e^{i(\epsilon_{\bf q}-\omega)t}(1+2n_{\bf q})\nonumber\\
&=&S\sqrt{2\pi}\delta(\epsilon_{\bf q}-\omega)\left(1+2\frac{1}{e^{\beta\omega}-1}\right).
\end{eqnarray}
Following essentially the same steps as the ones performed above, we obtain the
3D real space anticommutator for $\omega>\Delta_F$
\begin{eqnarray}
  S^\mathrm{3D}_{\perp}(\omega,{\bf q})&=&S\coth(\beta\omega/2)\times\\
&&\times \int_{-1}^{1}\int_{0}^{\infty}dxdq q^2 e^{iqrx}\delta(\epsilon_{{\bf q}}-\omega)\nonumber\\
&=&\frac{S}{D}\coth(\beta\omega/2)\frac{\sin(\sqrt{(\omega-\Delta_F)/D}r)}{r}\,.\nonumber\\
\end{eqnarray}
Let us now finally calculate the transverse susceptibility defined as
\begin{eqnarray}
\chi_\perp(t,{\bf q})&=&-i\theta(t)[S_{{\bf q}}^{+}(t),S_{-{\bf q}}^{-}(0)]\,.
\end{eqnarray}
As before, in the harmonic approximation, one finds
\begin{eqnarray}
\chi_\perp(t,{\bf q})&=&i\theta(t)2Se^{i\epsilon_{\bf q}t}\,.
\end{eqnarray}
In the frequency domain, we then have
\begin{align}
\chi_\perp(\omega,{\bf q})&=\frac{2iS}{\sqrt{2\pi}}\int_{0}^{\infty}dte^{i(\epsilon_{\bf
q}-\omega)t-\eta t}\\
&=-\frac{2S}{\sqrt{2\pi}}\frac{1}{\epsilon_{\bf
q}-\omega+i\eta}\nonumber\,,
\end{align}
and thus in the small ${\bf q}$ expansion
\begin{eqnarray}
\chi_\perp(\omega,{\bf q})&=&-\frac{2S}{\sqrt{2\pi}}\frac{1}{D{\bf q}^2+\Delta_F-\omega+i\eta}\,.
\end{eqnarray}
\begin{widetext}
In real space, for the three-dimensional case, we  obtain
\begin{eqnarray}
\chi^{\text {3D}}_\perp(\omega,{\bf r})&=&-\frac{2S}{\sqrt{2\pi}}\frac{2\pi}{(2\pi)^{3/2}}\int_{0}^{\infty}\int_{-1}^{1}dxdqq^2\frac{1}{D{\bf q}^2+\Delta_{F}-\omega+i\eta}e^{iqrx}\nonumber\\
&=&-\frac{4S}{\sqrt{2\pi}}\frac{2\pi}{(2\pi)^{3/2}}\frac{1}{r}\int_{0}^{\infty}dqq\frac{1}{Dq^2+\Delta_F-\omega+i\eta}\sin(qr)\,.
\end{eqnarray}
Making use of the Plemelj formula we obtain for $\omega>\Delta_F$
\begin{eqnarray}
\chi^{\text {3D}}_\perp(\omega,{\bf r})&=&
-\frac{2S}{\sqrt{2\pi}}\frac{2\pi}{(2\pi)^{3/2}}\frac{1}{r}\int_{-\infty}^{\infty}dq
q\frac{1}{Dq^2+\Delta_F-\omega+i\eta}\sin(qr)\nonumber\\
&=&-\frac{2S}{\sqrt{2\pi}}\frac{2\pi}{(2\pi)^{3/2}}\frac{1}{r}
P\int_{-\infty}^{\infty}dq
\frac{q}{Dq^2+\Delta_F-\omega}\sin(qr)+i\frac{2S}{\sqrt{2\pi}}\frac{2\pi^2}{(2\pi)^{3/2}}\frac{1}{r}\int_{-\infty}^{\infty}dq
q\delta(Dq^2+\Delta_F-\omega)\sin(qr)\nonumber\\
&=&-\frac{S}{D}\frac{\cos(r\sqrt{(\omega-\Delta_F)/D})}{r}+i\frac{S}{2D}\frac{\sin(\sqrt{(\omega-\Delta_F)/D}r)}{r}\,.
\end{eqnarray}
\end{widetext}
It is worth pointing out that the imaginary part of the susceptibility vanishes
\begin{equation}
\chi^{\text {3D}}_{\perp}(\omega,{\bf r})''=0,\quad \omega<\Delta_F,
\end{equation}
and therefore the susceptibility is purely real and takes the form of a Yukawa
potential
\begin{equation}
\chi_{\perp}^{\text{3D}}(\omega,{\bf r})=-\frac{S}{D}\frac{e^{-r/l_F}}{r},\quad\omega<\Delta_F,
\end{equation}
where $l_F=\sqrt{\frac{D}{\Delta_F-\omega}}$.

Note also that the imaginary part of the transverse susceptibility satisfies the
well-know fluctuation dissipation theorem
\begin{eqnarray}
  S^\mathrm{3D}_{\perp}(\omega,{\bf r})&=&\coth(\beta\omega/2)\chi^{\text {3D}}_\perp(\omega,{\bf r})^{\prime\prime}\,.
\end{eqnarray}

In three dimensions the susceptibility decays as $1/r$, where $r$ is measured in
lattice constants. For distances of order of $1\mu m$ this leads to a reduction by four orders
of magnitude.

For quasi one-dimensional ferromagnets such a reduction is absent and the
transverse susceptibility reads
\begin{equation}
  \chi_\perp^{\text{1D}}(\omega,r)=-\frac{S}{D}l_F e^{-r/l_F},\quad\omega<\Delta_F,
  \label{eq:chi_1D_r}
\end{equation}
where $l_F$ is defined as above and the imaginary part vanishes as above, i.e.,
\begin{equation}
\chi_{\perp}^{\text{1D}}(\omega,r)''=0,\quad\omega<\Delta_{F}.
\end{equation}
Similarly for $\omega>\Delta_F$ we have
\begin{equation}
\chi_{\perp}^{\text{1D}}(\omega,r)=S\frac{\sin\left(\sqrt{(\omega-\Delta_F)/D}r\right)}{\sqrt{D(\omega-\Delta_F)}}\,,
\end{equation}
and
\begin{equation}
\chi_{\perp}^{\text{1D}}(\omega,r)''=\frac{S}{2D}\sqrt{\frac{D}{\omega-\Delta_F}}\cos\left(\sqrt{(\omega-\Delta_F)/D}r\right)\,.
\label{eq:chi_perp1D}
\end{equation}

\section{Longitudinal correlators $\langle S_{{\bf q}}^{z}(t)S_{-{\bf q}}^{z}(0)\rangle$}
The longitudinal susceptibility reads
\begin{eqnarray}
\chi_\parallel(t,{\bf q})&=&-i\theta(t)[ S_{{\bf q}}^{z}(t),S_{-{\bf
q}}^{z}(0)]\\ &=&-\theta(t)\frac{1}{N}\sum_{{\bf q}',{\bf
q}''}e^{it(\epsilon_{{\bf q}'}-\epsilon_{{\bf q}'+{\bf q}})}\langle[a_{{\bf
q}'}^{\dagger}a_{{\bf q}'+{\bf q}},a_{{\bf q}''}^{\dagger}a_{{\bf q}''-{\bf
q}}]\rangle\nonumber\,.
\end{eqnarray}
Applying Wick's theorem and performing a Fourier transform, we obtain the
susceptibility in frequency domain
\begin{equation}
\chi_\parallel(\omega,{\bf q})=-\frac{1}{N}\sum_{{\bf k}}\frac{n_{\bf k}-n_{{\bf k}+{\bf q}}}{\omega-\epsilon_{{\bf k}+{\bf q}}+\epsilon_{\bf k}+i\eta}\,,
\label{eq:chi_par}
\end{equation}
where $n_{\bf k}$ is the magnons occupation number, which is given by the
Bose-Einstein distribution 
\begin{equation}
n_k=\frac{1}{e^{\beta \epsilon_{\bf k}}-1}\,,
\end{equation}
where $\epsilon_{\bf k}$ is again the magnon spectrum ($\epsilon_{\bf
k}=\omega_{\bf k}+\Delta_F\approx D{\bf k}^2+\Delta_F$ for small $k$). Note that
the transverse susceptibility dominates, since  its ratio to longitudinal one
is proportional to $1/S\ll 1$. Furthermore, the longitudinal susceptibility is
$\propto n_k \ll 1$ and thus vanishes at zero temperature, while this is not
the case for the transverse.
 
Since we are interested in the decoherence processes caused by the longitudinal
fluctuations, we calculate the imaginary part of $\chi_\parallel(\omega,{\bf
q})$ which is related to the fluctuations via the
\textit{fluctuation-dissipation theorem}. Performing a small ${\bf q}$
expansion and assuming without loss of generality $\omega>0$, we obtain
\begin{widetext}
\begin{eqnarray}
\chi^{\text{3D}}_\parallel(\omega,{\bf q})^{\prime\prime}&=&
\frac{\pi}{(2\pi)^3}\int d{\bf
k}(n_{\bf k}-n_{{\bf k}+{\bf q}})\delta(\omega_{\bf k}-\omega_{{\bf k}+{\bf
q}}+\omega)\nonumber\\
&=&\frac{1}{4\pi}\int_{0}^{\infty}dk
k^2\int_{-1}^{1}dx\left(\frac{1}{e^{\beta(\Delta_F+ D k^2)}-1}-\frac{1}{e^{\beta
(\omega+\Delta_F+Dk^2)}-1}\right)\delta(\omega-D q^2-2Dkqx)\nonumber\\
&=&\frac{1}{4\pi}\int_{0}^{\infty}dk
k^2\int_{-1}^{1}dx\left(\frac{1}{e^{\beta(\Delta_F+D k^2)}-1}-\frac{1}{e^{\beta
(\omega+\Delta_F+Dk^2)}-1}\right)\delta\left(k-\frac{\omega-Dq^2}{2Dqx}\right)\left\vert\frac{1}{2Dqx}\right\vert\nonumber\\
&=&\frac{1}{4\pi}
\int_{-1}^{1}dx\left\vert\frac{1}{2Dqx}\right\vert\left(\frac{\omega-Dq^2}{2Dqx}\right)^2
\left(\frac{1}{e^{\beta
  \left(\Delta_F+D\left(\frac{\omega-Dq^2}{2Dqx}\right)^2\right)}-1}-\frac{1}{e^{\beta
    \left(\omega+\Delta_F+D\left(\frac{\omega-Dq^2}{2Dqx}\right)^2\right)}-1}\right)\theta\left(\frac{\omega-Dq^2}{2Dqx}\right)\nonumber\\
    &=&\frac{1}{4\pi}\int_{0}^{1}dx\frac{1}{2Dqx}\left(\frac{\omega-Dq^2}{2Dqx}\right)^2
\left(\frac{1}{e^{\beta
  \left(\Delta_F+D\left(\frac{\omega-Dq^2}{2Dqx}\right)^2\right)}-1}-\frac{1}{e^{\beta
    \left(\omega+\Delta_F+D\left(\frac{\omega-Dq^2}{2Dqx}\right)^2\right)}-1}\right).
  \end{eqnarray}
\end{widetext}
Next, since we are interested in the regime where $\omega\gg T$ (and thus
$\beta\omega\gg 1$), we have $n_{\bm k}\gg n_{\bm{k+q}}$. Furthermore, we approximate
the distribution function $n_{\bm k}=\frac{e^{-\beta(\Delta_F+\omega_{\bm
k})}}{1-e^{-\beta\Delta_F}+\beta\omega_{\bm k}}$ (this is valid when
$\beta\omega_{\bf k}\ll 1$) and arrive at the following expression
\begin{widetext}
\begin{align}
    \label{eq:qzz}
    \chi^{\text{3D}}_\parallel(\omega,{\bf q})^{\prime\prime}&=\frac{1}{4\pi}\int_{0}^{1}dx\frac{1}{2Dqx}\left(
    \frac{\omega-Dq^2}{2Dqx} \right)^2 \frac{e^{-\beta\left(\Delta_F+D\left(
      \frac{\omega-Dq^2}{2Dqx} \right)^2\right)}}{1-e^{-\beta\Delta_F}+\beta D\left(
    \frac{\omega-Dq^2}{2Dqx} \right)^2 }\nonumber\\
    &=-\frac{e^{1-e^{-\beta\Delta_F}-\beta\Delta_F}}{4\beta D^2q}\mathrm {Ei}\left(e^{-\beta\Delta_F}+\frac{1}{4}\left(-4-\beta
    Dq^2+2\beta\omega-\frac{\beta\omega^2}{Dq^2}\right)\right)\,,
\end{align}
\end{widetext}
where $\mathrm{Ei}(z)$ is the exponential integral function. We also need the
the real space represenation obtained after inverse Fourier transformation,
\begin{align}
  \chi^{\text{3D}}_\parallel(\omega,{\bf r})^{\prime\prime}&=\sqrt{\frac{2}{\pi}}\frac{1}{r}\int_{0}^{\infty}dqq
\chi^{\text{3D}}_\parallel(\omega,q)^{\prime\prime} \sin(qr)\,.
\label{eq:chi_zzr}
\end{align}

In order to perform the above integral we note that the imaginary part of the
longitudinal susceptibility, given by Eq.~(\ref{eq:qzz}), is peaked
around $q=\sqrt{\omega/D}$ with the width of the peak ($1/\sqrt{\beta D}$)
much smaller than its position in the regime we are working in ($\omega\gg T$).
For $\bf r=0$, the integration over $q$ can be then performed approximately and
yields the following expression

\begin{align}
  \chi^{\text{3D}}_\parallel(\omega,{\bf r=0})^{\prime\prime}&=\frac{\sqrt{\pi}e^{-e^{-\beta\Delta_F}-3\beta\Delta_F/2}}{2\beta^2D^3}
  \left( e^{e^{-\beta\Delta_F}+\beta\Delta_F/2}\right.\nonumber\\
  &-e\sqrt{\pi}\sqrt{e^{\beta\Delta_F}-1}\\
  &\times\left.\mathrm{Erfc}(e^{-\beta\Delta_F/2}\sqrt{e^{\beta\Delta_F}-1})\right)\sqrt{\beta\omega}\,,\nonumber
\end{align}
where $\mathrm{Erfc}(z)$ denotes the complementary error function. It is
readily observed from the above expression that the longitudinal fluctuations
are exponentially suppressed by the gap. Assuming that $\Delta_F\gg T$, we
obtain the following simplified expression
\begin{align}
  \chi^{\text{3D}}_\parallel(\omega,{\bf r=0})^{\prime\prime}&=\frac{\sqrt{\pi}-e\pi\mathrm{Erfc}(1)}{2\beta^2D^3}e^{-\beta\Delta_F}\sqrt{\beta\omega}\,.
  \label{eq:spectraldensity_longitudinal}
\end{align}
We observe that, since $J(\omega)=\chi_\parallel(\omega,{\bf
r})^{\prime\prime}$, the longitudinal noise of the ferromagnet is---as the
transverse one---sub-ohmic~\cite{divincenzo_rigorous_2005}.

Next we calculate the longitudinal fluctuations for the case of a quasi-
one-dimensional ferromagnet ($\Delta_F\gg T$) and obtain

\begin{align}
  \chi_\parallel^{\text{1D}}(\omega,r=0)^{\prime\prime}&=\frac{1}{4\pi}\int_{-\infty}^{\infty}dk\int_{-\infty}^{\infty}dq
\left(\frac{1}{e^{\beta(\Delta_F+ D k^2)}-1}\right.\nonumber\\
&-\left.\frac{1}{e^{\beta (\omega+\Delta_F+Dk^2)}-1}\right)\delta(\omega-D
q^2-2Dkq)\nonumber\\
&=\int_{-\infty}^\infty dk \frac{e^{-\beta Dk^2}}{1-e^{-\beta\Delta_F}+\beta
Dk^2}\frac{1}{D\sqrt{k^2+\omega/D}}\nonumber\\
&=\frac{\gamma}{D\sqrt{\beta\omega}}e^{-\beta\Delta_F}\,,
  \label{eq:fluct_zz_1D}
\end{align}
where $\gamma$ is a numerical factor of order 1.

\section{Rotated Hamiltonian for CNOT Hate}
\label{sec:rotated}
\begin{widetext}
Here we give the general for of the matrix $\hat{\widetilde{H}}$ entering Eq.~(\ref{eq:Heffgentilde}).
\begin{align}
\tilde{\mathcal H}_{12}&=\frac{3\left(\Delta_z \Delta_z^- (C_2^{\prime\prime} (-6 \Delta_x A_1^\prime+B_1 \Delta_z^-+3 C_1^\prime \Delta_z^+)+2 \Delta_x A_1^{\prime\prime} (B_2+3 C_2^\prime)+C_1^{\prime\prime} (B_2 \Delta_z^--3 C_2^\prime \Delta_z^+))\right)}{32 \Delta ^3}\\
  &+\frac{3\Delta_x i \left(12 A_2 \Delta  \Delta_x A_1^*-12 A_1 \Delta  \Delta_x A_2^*+\Delta_x \left(-4 \Delta_x A_1^\prime (B_2+3 C_2)+4 A_1 B_2 \Delta_x+\Delta_z^- \left(3 C_1 C_2^*+2 i (B_1 C_2^{\prime\prime}+B_2 C_1^{\prime\prime})\right)\right)\right)}{64\Delta^3}\nonumber\\
  &+\frac{3\Delta_xi\left(12 A_1 \Delta_x^2 C_2^\prime+3 C_1^* \Delta_z^- (2 A_2 \Delta -C_2 \Delta_x)+\left(2 B_1 \Delta  A_2^{\prime\prime} \Delta_z^-+3 i C_1 \Delta  A_2^* \Delta_z^-\right)\right)}{64\Delta^3}\,,\nonumber\\
\tilde{\mathcal H}_{13}&=\frac{2 \left(\Delta_z^z (B_2 \Delta_x (B_1-3 C_1^\prime)-3 B_1 (2 \Delta_z A_2^\prime+\Delta_x C_2^\prime))-6 B_2 \Delta_x^2 A_1^\prime\right)+18 \Delta_z A_2^* (2 A_1 \Delta_x+C_1 \Delta -C_1 \Delta_z)}{{64 \Delta ^2}}\\
  &+\frac{9 \Delta_x C_2^* (2 A_1 \Delta_x+C_1 \Delta -C_1 \Delta_z)+9 C_1^* \Delta_z^- (2 A_2 \Delta_z+C_2 \Delta_x)+18 \Delta_x A_1^* (2 A_2 \Delta_z+C_2 \Delta_x)}{64 \Delta ^2},\nonumber\\
\tilde{\mathcal H}_{23}&=\frac{3 i\left(2 i (\Delta_x( \Delta_z^- (B_1 C_2^{\prime\prime}+B_2 C_1^{\prime\prime})+2 i B_2 \Delta_x A_1^\prime)+2 i B_1 \Delta_z A_2^\prime \Delta_z^-)+4 \left(A_1 B_2 \Delta_x^2+A_2 B_1 \Delta_z \Delta_z^-\right)\right)}{64\Delta^2}\\
  &+\frac{3i\left(6 \Delta_z A_2^* (-2 A_1 \Delta_x-C_1 \Delta +C_1 \Delta_z)-6 \Delta_x A_1^* (C_2 \Delta_x-2 A_2 \Delta_z)+3 \Delta_x C_2^* (2 A_1 \Delta_x-C_1 \Delta +C_1 \Delta_z)\right)}{64\Delta^2}\nonumber\\
  &+\frac{3i\left(3 C_1^* \Delta_z^- (2 A_2 \Delta_z+C_2 \Delta_x)\right)}{64 \Delta ^2},\nonumber
\end{align}
\begin{align}
\tilde{\mathcal H}_{11}&=\frac{2 (6 B_2 \Delta_x \Delta_z A_1^\prime-\Delta_z^- (6 B_1 \Delta_x A_2^\prime+B_2 \Delta_z (B_1-3 C_1^\prime)-3 B_1 \Delta_z C_2^\prime))+18 \Delta_x A_2^* (2 A_1 \Delta_x+C_1 \Delta -C_1 \Delta_z)}{32 \Delta ^2}\nonumber\\
  &+\frac{18 \Delta_x A_1^* (2 A_2 \Delta_x-C_2 \Delta_z)+9 \Delta_z C_2^* (-2
  A_1 \Delta_x-C_1 \Delta +C_1 \Delta_z)-9 C_1^* \Delta_z^- (C_2 \Delta_z-2 A_2
  \Delta_x)}{32 \Delta ^2},\\
  \tilde{\mathcal H}_{22}&=\frac{2 (6 B_2 \Delta_x A_1^\prime+\Delta_z^- (B_1 (B_2+3 C_2^\prime)+3 B_2 C_1^\prime))-9 C_2^* (2 A_1 \Delta_x-C_1 \Delta +C_1 \Delta_z)-18 C_2 \Delta_x A_1^*+9 C_2 C_1^* \Delta_z^-}{32 \Delta },\\
\tilde{\mathcal H}_{33}&=0,
  \label{eq:matrixelH}
\end{align}
and the rest of the components $\tilde{\mathcal H}_{ij}$ are obtain from
$\tilde{\mathcal H}_{ji}$ by exchanging $i\leftrightarrow j$.
\end{widetext}

\bibliography{ref}
\end{document}